\documentclass[12pt, full]{article}
\usepackage{amsthm}
\usepackage{graphicx} % Required for inserting images
\usepackage{amsmath}
\usepackage{dirtytalk}
\usepackage{booktabs}
\usepackage{rotating}
\usepackage{array}
\usepackage{float}
\usepackage{bbm}
\theoremstyle{plain} 
\usepackage{pdfpages}
\newtheorem{theorem}{Theorem}
\newtheorem{assumption}{Assumption}

\newtheorem{proposition}{Proposition}

\newtheorem{corollary}{Corollary}

\usepackage{caption} 
\usepackage{subcaption}

\theoremstyle{remark} % alternative style, e.g. for remarks
\newtheorem{remark}{Remark}
\usepackage{ragged2e}
 \usepackage[T1]{fontenc}
\usepackage[utf8]{inputenc}
\usepackage{newtxmath}
 \usepackage{stix2}

\usepackage{geometry}
\usepackage{xcolor}
\definecolor{blue}{RGB}{0,114,178}

\usepackage{natbib}
\usepackage{hyperref}
\hypersetup{
  colorlinks = true,
  linkcolor = blue,
  citecolor = blue,
  urlcolor = black
}

\usepackage[toc,page]{appendix}

\title{Microfoundations and the Causal Interpretation of Price-Exposure Designs}
\author{%
\begin{tabular*}{\textwidth}{@{\extracolsep{\fill}}ccc}
Luca Moreno-Louzada & Guilherme Figueira & Pedro Picchetti\\
\shortstack{\textit{\small Stanford}\\[-0.1em]\textit{\small University}} &
\shortstack{\textit{\small São Paulo}\\[-0.1em]\textit{\small School of Economics}} &
\shortstack{\textit{\small Pontificia Universidad }\\[-0.1em]\textit{\small Católica de Chile}}
\end{tabular*}
}
\date{Work in progress\\ This version: December 2025}

\date{\small Work in progress\\ This version: December 2025}

\begin{document}

\maketitle
\renewcommand\thefootnote{} % Remove numbering for the next footnote
\footnote{ \\ We thank Luis Alvarez, Bruno Ferman, Matthew Gentzkow, Isaac Sorkin, and Laura Weiwu for helpful feedback and suggestions. This is a work in progress; any comments are welcome. Emails: lucamlouzada@gmail.com; guilhermefigueira@live.com; pedropicchetti@gmail.com }
\renewcommand\thefootnote{\arabic{footnote}} % Restore normal numbering
\vspace{-1.5cm}
\begin{abstract}
This paper studies regional exposure designs that use commodity prices as instruments to study local effects of aggregate shocks. Unlike standard shift–share designs that leverage differential exposure to many shocks, the price–exposure relies on exogenous variation from a single shock, leading to challenges for both identification and inference. We motivate the design using a multi-sector labor model. Under the model and a potential outcomes framework, we characterize the 2SLS and TWFE estimands as weighted averages of region- and sector-specific effects plus contamination terms driven by the covariance structure of prices and by general-equilibrium output responses. We derive conditions under which these estimands have a clear causal interpretation and provide simple sensitivity analysis procedures for violations. Finally, we show that standard inference procedures suffer from an overrejection problem in price-exposure designs. We derive a new standard error estimator and show its desirable finite-sample properties through Monte Carlo simulations. In an application to gold mining and homicides in the Amazon, the price–exposure standard errors are roughly twice as large as conventional clustered standard errors, making the main effect statistically insignificant.

\end{abstract}

% \noindent\textbf{Keywords:} Price‐exposure designs; commodity price shocks; instrumental variables; two‐way fixed effects; shift‐share identification; continuous‐treatment DiD; heterogeneous treatment effects; labor‐market elasticities.\\
% \noindent\textbf{JEL classification:} C21, C23, C26, C33, F16

\section{Introduction}
A pervasive empirical strategy in economics is to use regional-exposure designs with panel data to exploit heterogeneous exposure across locations facing a common shock.\footnote{See \citet{borusyak2023nonrandom} and \citet{karthikexposuredesign}} This paper studies a particularly influential class of these strategies that leverages international commodity prices to identify local causal effects of a specific economic activity. The key empirical variable is typically constructed as:

\begin{equation*}
 Z_{it} = A_{i} \times P_t 
\end{equation*}

 where $ A_i $ denotes regional exposure which is fixed in time (e.g., mineral endowments), and $ P_t $ is a time-varying aggregate price which is the same for all units (e.g., the log international gold price). Unlike the typical shift-share design, which aggregates several shocks using weights for differential exposure, the \textit{price-exposure design} focuses on a single shock.

This setup has been used, for example, to study the effect of different commodities on conflict \citep{dube2013commodity, berman2017mine} and of cocaine production on child outcomes \citep{sviatschi2022making}.\footnote{Other applications span the effects of oil exploitation on public finances \citep{martinez2023natural}, health spending \citep{acemoglu2013income}, crime \citep{soares2025too}, human capital \citep{balza2025local}, and political outcomes \citep{carreri2017natural, fetzer2024cohesive}; coffee production on schooling \citep{carrillo2020present}; maize on drug production \citep{dube2016maize}; and mining on human capital \citep{mejia2020mining}, crime \citep{vasquez2025peer}, household consumption \citep{bazillier2020gold}, deforestation \citep{girard2025artisanal}, and proto-state formation \citep{sanchez2020origins}.} However, while this approach is ubiquitous, there is little formal discussion about its underlying identification assumptions and econometric properties. In particular, it is often unclear what causal parameter is recovered in a two-stage least squares (2SLS) or two-way-fixed-effects (TWFE) regressions, what assumptions about the first stage justify its causal interpretation, and how local labor-market dynamics or general equilibrium responses affect identification.

The first contribution of this paper is to provide microfoundations for the price–exposure design using a stylized multi-sector labor model. In our model, changes in the world price of a commodity in the sector of interest (we denote it as sector $q$) affect local wages and the allocation of labor across the other sectors, which determines the first stage relationship even for outcomes that are not themselves labor-market variables (e.g., human capital, crime, or conflict). This stylized environment makes explicit that a regression of $Y_{it}$ on $Z_{it}$ (or of $Y_{it}$ on $X_{it}$ instrumented with $Z_{it}$) implicitly imposes strong restrictions on region-specific labor supply elasticities, sectoral demand parameters, and the way in which prices in different sectors move together. We derive closed-form expressions for the sector-$q$ first stage and show that the slope of the relationship between $Z_{it}$ and sector-$q$ output is heterogeneous across regions and can even change sign when local labor supply is inelastic and exposure to co-moving sectors is high. This yields a transparent monotonicity condition for price–exposure designs.

Second, we place price–exposure regressions into a potential-outcomes framework and characterize the finite-population estimands of common estimators. We follow \citet{adao2019shift} --- hereafter AKM --- and treat the observed panel of regions and periods as the population of interest, with uncertainty coming from the stochastic process for prices. Under a linear potential-outcomes model in sectoral outputs, we show that the 2SLS estimand obtained by instrumenting sector-$q$ output $X_{iqt}$ with $Z_{iqt} = A_{iq} p_{qt}$ can be decomposed into two terms: (i) a weighted average of region-specific causal effects from an increase in the output of sector $q$, with weights proportional to the elasticity in the first-stage, exposure and the variance of the price in sector $q$; and (ii) a contamination term that aggregates effects of other sectors $\beta_{is}$ weighted by the elasticity from the other sectors, exposures, and the covariance of prices from other sectors with the price in sector $q$.

The IV estimand is therefore a convex average of sector-$q$ effects only when (a) the first stage is monotone (the price-elasticity of the output in sector $q$ is nonnegative for all regions) and (b) the focal price is uncorrelated with other prices. Otherwise, it is a “weakly causal’’ object in the sense of \citet{blandhol2022tsls}, combining multiple channels in a model-dependent way. A similar contamination term appears even in the absence of price co-movement if the potential outcomes depend on the output of multiple sectors and general equilibrium responses lead to changes in sector-$q$ price affecting outputs in other sectors.

Third, we clarify the interpretation of TWFE specifications that regress $Y_{it}$ on $Z_{iqt}=A_{iq} p_{qt}$ with region and time fixed effects, a specification often described as a continuous-treatment differences-in-differences (DiD) or a reduced-form intention-to-treat (ITT). We show that the TWFE estimand can also be written as a weighted average of the same region-specific $\beta_{iq}$, plus a contamination term involving other sectors, but with weights that depend only on the variance and covariance of changes in prices. In contrast to IV, the TWFE estimand can be given a weakly causal decomposition under a 
weaker exogeneity assumption: mean-independence of $\Delta p_{qt}$ from changes in 
potential outcomes and sectoral outputs, conditional on the finite-population filtration 
of effects and first stage elasticities. This assumption resembles a parallel–trends restriction in first differences, but it operates on the common price shocks rather than on outcome trends. Our derivation highlights that this design is not the same as difference-in-differences and hence does not require the same identification assumptions. Under additional restrictions on the first stage (for instance, 
homogeneous elasticities equal to 1 for all regions), together with 
monotonicity and absence of cross-price comovement, the TWFE estimand coincides with a 
convex weighted average of sector-$q$ effects. We revisit the reduced form specifications in \citet{dube2013commodity} and \citet{sviatschi2022making} to discuss how including controls for other price shocks can affect the magnitudes of estimated effects.

Building on these decompositions, we derive sensitivity formulas that quantify how much the 2SLS estimand can be shifted by plausible values of the contamination term. We show how assumptions about the magnitude of the contamination term map into an identified set for the weighted average of interest, and we highlight “breakdown’’ values at which qualitative conclusions (e.g., the sign of the effect) change. 

Finally, we study estimation and inference for price–exposure designs. Within the finite–population framework, we show the 2SLS estimator based on $Z_{iqt}=A_{iq}p_{qt}$ is consistent and asymptotically normal at rate $\sqrt{NT}$, with an asymptotic variance that takes a shift–share–type randomization form analogous to AKM, but exploiting time–series rather than cross–sectional variation. We use this characterization to construct a feasible ``price–exposure'' (PE) variance estimator that aggregates residuals across regions using only the variation in $p_{qt}$, and we show that it delivers valid inference under heterogeneous effects and arbitrary correlation in the outcome shocks. In contrast, conventional Eicker–Huber–White standard errors can severely understate sampling uncertainty when residuals inherit a shift–share structure from other prices, leading to systematic overrejection. 

We document the magnitude of these distortions in Monte Carlo simulations under different combinations of $N$ and $T$, and show that the PE standard errors achieve close–to–nominal coverage once the panel is moderately rich in the time dimension. To illustrate the practical relevance of these issues, we carry out a stylized empirical application to gold mining and homicides in the Brazilian Amazon. In this example, conventional clustered standard errors suggest a precisely estimated positive effect, whereas the PE standard errors are roughly twice as large and render the estimate statistically insignificant, highlighting how it can be important to take the specific sampling structure of price–exposure designs into account in empirical settings.

We conclude with brief guidance for applied work, emphasizing the importance of transparent discussion of the potential outcomes and identifying assumptions underlying causal claims. At a minimum, empirical researchers using price–exposure designs should state explicitly how the price process enters the relevant potential outcomes in their setting. Our goal is to offer a framework that clarifies the target parameters in price–exposure designs and serves as a starting point for these discussions.

Our analysis relates to long-standing questions in econometrics about the interpretation of causal parameters. \citet{angrist2000interpretation} showed that, under standard IV assumptions, the 2SLS estimator identifies a weighted average of marginal treatment effects. Recent work has extended these insights to characterizes how these weights depend on treatment effect heterogeneity \citep{blandhol2022tsls,alvarez2024interpretation}. A parallel literature studies difference-in-differences and two-way fixed-effects estimands as weighted averages of underlying causal effects, documenting the possibility of negative or counterintuitive weights in designs with a continuous treatment \citep{de2018fuzzy,de2024difference, callaway2024difference}. We build on these perspectives by showing that price-exposure estimands can be written as weighted averages of region- and sector-specific causal effects, with weights that have a transparent economic interpretation, and by characterizing when these weights are convex. 

A second strand of the literature that we contribute to studies exposure designs, in which a common aggregate shock is transmitted through heterogeneous exposure shares \citep[e.g.,][]{goldsmith2020bartik, borusyak2023nonrandom}. Within this literature, we focus on the special case in which the aggregate shock is a single commodity price and develop a microfoundation and design-based inference strategy tailored to this setting. We build on results on shift-share identification with exogenous shocks \citep{adao2019shift,borusyak2022quasi}, emphasizing a many-time-period rather than many-sector asymptotic framework, and on work showing how inference is complicated when exposure instruments load on a single aggregate shock \citep{arkhangelsky2019policy,karthikexposuredesign}, exploiting the structure of prices to microfound and interpret the parameters of interest.

The rest of the article is organized as follows. We begin by discussing motivating examples of how the design has been implemented in previous empirical work in Section \ref{sec:applied_literature}. Section \ref{sec:economic_model} presents the economic model which microfounds the design. Section \ref{sec:potential_outcomes} develops a potential-outcomes framework and studies the causal interpretation of IV and TWFE estimands. Section \ref{sec:estimation_inference} discusses estimation and inference. Section \ref{sec:montecarlo} reports Monte Carlo simulations. Section \ref{sec:empirical_examples} discusses examples from empirical applications, and Section \ref{sec:conclusion} concludes.

\subsection{Price Exposure Designs in the Applied Literature}
\label{sec:applied_literature}
Applied researchers are often interested in estimating the causal effect of some economic activity on an outcome. Price–exposure designs proxy for endogenous or unobserved local activity by interacting a time-invariant exposure $A_i$ (e.g., geological endowments, suitability, historic production) with an aggregate price $P_t$ common to all regions, building $Z_{it}=A_i\times P_t$. Two implementations are typical: (i) a reduced-form TWFE regression of outcomes on $Z_{it}$, interpreted as an ITT effect, and often called a variant of DiD; and (ii) an IV specification that instruments a measure of local activity $X_{it}$ with $Z_{it}$. Both approaches implicitly assume a positive and sufficiently homogeneous first stage linking $Z_{it}$ to $X_{it}$, an assumption rarely discussed in detail. 

Below, we summarize a few examples of how this design has been used in applied work.\footnote{We simplify and adapt the original notation of the papers to standardize methodologies across different settings.}

\paragraph{\citet{dube2013commodity}} The authors study the effects of commodity price shocks on violence in Colombia. Using a municipality-year panel with data from 1988 to 2005, they construct two price‐exposure variables, $Z^{\text{oil}}_{it}=q^{\text{oil}}_{i,1988}\times p^{\text{oil}}_{t}$ and $Z^{\text{coffee}}_{it}=q^{\text{coffee}}_{i,1997}\times p^{\text{coffee}}_{t}$. Here $q^{\text{oil}}_{i,0}$ is used as the pre‐period production, and $Z^{\text{coffee}}_{mt}$ is instrumented with an interaction of local temperature and rainfall and production in other countries. They estimate it with 2SLS as a single equation including both oil and coffee, and call it a difference-in-differences estimator where time variation stems from movements in annual prices. 

% They use both commodities to differentiate between the positive \say{opportunity cost} effects of income shocks in labor-intensive activities (such as coffee), and the negative \say{rapacity} effect of non-labor intensive activities (such as oil), in which increases in income lead to more violence by raising returns to appropriation. Though their main analysis contrasts oil and coffee, they also run specifications with other commodities. 

\paragraph{\citet{berman2017mine}} 
The study analyzes $55\text{ km}\times55\text{ km}$ grid cells across Africa, defining treatment as the interaction $Z_{it}=\mathbb{1}\{\text{number of mines}_{i}> 0\}\times p^{m}_t$, where $p^{m}_t$ is the international price of the mineral extracted in cell $i$. The authors estimate different specifications with time and cell-fixed effects, and  do not elaborate on what the strategy is called. They find that conflict increases when the price rises. 

\paragraph{\citet{carrillo2020present}} 
The paper exploits municipality‐level heterogeneity in pre-existing coffee cultivation to build the exposure term $Z_{it}=q^{\text{coffee}}_{i,0}\times p^{\text{coffee}}_{t}$, where $q^{\text{coffee}}_{i,0}$ is baseline coffee area and $p^{\text{coffee}}_{t}$ is the world price of coffee. A cohort-specific reduced form specification compares children whose school-age years coincided with coffee booms to those who faced busts, controlling for municipality and birth‐cohort fixed effects. The paper finds that higher coffee prices lower completed years of schooling and depress adult earnings.

\paragraph{\citet{sviatschi2022making}} 
The author studies Peruvian districts whose geographical conditions make them suitable for coca. Treatment is $Z_{it}=s^{\text{coca}}_{i}\times \widehat{p^{\text{coca}}_{t}}$, with $s^{\text{coca}}_{i}$ fixed suitability and $\widehat{p^{\text{coca}}_{t}}$ the price of coca leaves in Peru instrumented by production in Colombia. A TWFE specification shows that price spikes raise child labor in coca farming, reduce schooling, and increase the probability of incarceration three decades later. 

\paragraph{\citet{mejia2020mining}} 
This work combines satellite-based measures of gold mining pits with Colombian administrative school records and defines exposure as $Z_{it}=s_{i}^{\text{gold}}\times p^{\text{gold}}_{t}$. where $s_i$ denotes the standardized number of gold deposits in the vicinity of school $i$. Using school and year fixed effects, the author shows both reduced form TWFE estimates and 2SLS where the number of cells with detected mining activity is instrumented by $Z_{it}$. The paper finds that mining raises primary enrollment and lowers dropout, but harms test scores and tertiary enrollment. 

\paragraph{\citet{balza2025local}} 
The authors assemble well-level drilling data and micro-school enrollment records for Colombia. Their instrument is $Z_{it}=s^{\text{oil}}_{i}\times p^{\text{oil}}_{t}$, where $s^{\text{oil}}_{i}$ is a proxy for local oil endowment $i$ and $p^{\text{oil}}_{t}$ is the global oil price. A 2SLS specification, with the first stage predicting the number of wells drilled, shows that oil exploitation reduces higher-education enrollment and delays college entry.

\section{Economic Model}
\label{sec:economic_model}

In this section, our goal is to derive a microfoundation for the first stage relationship between $Z_{it}$, the price-exposure instrument, and $X_{it}$, the local activity that researchers are interested in. We develop a simple multi-sector model with region-specific labor to illustrate how omitting explicit equilibrium conditions can complicate interpretation of estimated coefficients due to heterogeneous elasticities between regions, general equilibrium responses, and co-movements in commodity prices. The model draws heavily from Appendix C in AKM. 

\paragraph{Regions and sectors.}
Consider an economy composed of multiple regions, indexed by $i=1,\dots,N$.
Each region $i$ is endowed with a fixed labor supply $\bar{L}_i$, which is fully employed and can be reallocated \emph{across sectors} within a region but is \emph{immobile across regions}. Sectors are indexed by $s\in\mathcal S\equiv \mathcal S_T\cup\{0\}$, where $\mathcal S_T$ are tradable sectors with exogenous world prices $P_{st}$ and $s=0$ is a nontradable (internal) sector whose price is normalized to $1$. Throughout, it is useful to think of a tradable sector as a sector producing a commodity such as gold; $P_{gold, t}$ is then the international price of gold at time $t$.

\paragraph{Production.}
In tradable sector $s\in\mathcal S_T$ and region $i$, output at time $t$ is
\begin{equation}
\label{eq:prod-CD}
X_{ist} \;=\; A_{is}\,E_{ist}\,K_{is}^{\,1-\theta_s}\,L_{ist}^{\,\theta_s},
\qquad 0<\theta_s<1,
\end{equation}
where $A_{is}>0$ is a time-invariant productivity/capacity shifter (e.g.\ the number of mineral gold deposits in our running example),
$E_{ist}>0$ is a time-varying sector-region productivity shock,
$K_{is}>0$ is a sector-region capital stock (time-invariant within the short-run window),
and $L_{ist}$ is labor in sector $s$.
The internal sector produces
\begin{equation*}
I_{it} \;=\; B_{i}\,V_{it}\,K_{i0}^{\,1-\theta_0}\,N_{it}^{\,\theta_0},\qquad 0<\theta_0<1.
\end{equation*}
with notation analogous to \eqref{eq:prod-CD} and labor $N_{it}$.

\paragraph{Preferences/demand and market structure.}
Within each sector $s$, a standard CES aggregator across varieties with elasticity of substitution $\sigma_s>1$ yields the usual isoelastic revenue schedule; firms are price takers up to a constant markup (which is absorbed into sectoral constants).

\paragraph{Wage equalization and labor supply.}
Within region $i$, labor is mobile across sectors, so a single wage $w_{it}$ clears markets.
Reduced-form labor supply in region $i$ is
\begin{equation}
\label{eq:supply}
L_{it}^{\,S} \;=\; v_{it}\,w_{it}^{\,\phi_i},\qquad \phi_i>0,
\end{equation}
where $v_{it}$ is a (possibly slowly varying) shifter capturing demographics, participation, etc.
Total labor demand is $L_{it}^{\,D} \equiv \sum_{s\in\mathcal S_T}L_{ist}+N_{it}$.
Equilibrium requires $L_{it}^{\,S}=L_{it}^{\,D}$.

\paragraph{Sectoral Labor Demand}

Profit maximization in \eqref{eq:prod-CD} with CES demand implies,
for each tradable $s\in\mathcal S_T$,
\begin{equation}
\label{eq:Ld-AKM}
L_{ist}
\;=\;
\psi_{is}\;
w_{it}^{-\sigma_s\delta_s}\;
\bigl(A_{is}E_{ist}P_{st}\bigr)^{(\sigma_s-1)\delta_s}
\;K_{is}^{\,(1-\theta_s)(\sigma_s-1)\delta_s},
\end{equation}
where $\psi_{is}>0$ is a time-invariant constant that collects markups and sectoral demand primitives, and $\delta_s \equiv \frac{1}{1 + (\sigma_s - 1)(1-\theta_s)}$. $\delta_s\in(0,1)$ is a reduced-form parameter governing the sensitivity of sector-$s$ labor demand to the revenue shifter $A_{is}E_{ist}P_{st}$, which depends on technology through $\theta_s$ and on the elasticity of substitution $\sigma_s$.

Taking logs and deviations from a reference equilibrium $(\cdot)^*$,
\begin{equation}
\label{eq:dlnL_is_general}
d\ln L_{ist}
\;=\;
(\sigma_s-1)\delta_s\,d\ln P_{st}
\;+\;
(\sigma_s-1)\delta_s\,d\ln E_{ist}
\;-\;
\sigma_s\delta_s\,d\ln w_{it},
\end{equation}
where all other terms in \eqref{eq:Ld-AKM} are time-invariant and thus drop out in first differences.\footnote{If capital $K_{is}$ varies at horizon $t$, an additional term
$(\sigma_s-1)\delta_s(1-\theta_s)\,d\ln K_{is}$ appears; we treat $K_{is}$ as fixed over the short run.}

For the internal sector $s=0$ (price $1$) we obtain the same form with $P_{0t}\equiv 1$ and $E_{i0t}\equiv V_{it}$:
\begin{equation}
\label{eq:dlnL_i0}
d\ln N_{it}
\;=\;
(\sigma_0-1)\delta_0\,d\ln V_{it}
\;-\;
\sigma_0\delta_0\,d\ln w_{it}.
\end{equation}

\paragraph{Normalization and Baselines}

Let $\ell_{is,0}^*\equiv L_{is}^*/L_i^*$ denote baseline employment shares (with $s=0$ covering the internal sector). Fix baselines $\bar P_s,\bar E_{is},\bar v_i$ and define centered shocks
\[
p_{st}\equiv \ln P_{st}-\ln\bar P_s,\qquad
e_{ist}\equiv \ln E_{ist}-\ln\bar E_{is},\qquad
\nu_{it}\equiv \ln v_{it}-\ln\bar v_i,
\]
with $\mathbb E_t[p_{st}]=\mathbb E_t[e_{ist}]=\mathbb E_t[\nu_{it}]=0$.

\paragraph{Equilibrium Wage Feedback}

With the normalizations above, linearizing the market-clearing condition $L_{it}^{\,S}=L_{it}^{\,D}$ using \eqref{eq:supply}, \eqref{eq:dlnL_is_general}, and \eqref{eq:dlnL_i0} yields
\begin{align}
\phi_i\,d\ln w_{it} + d\ln v_{it}
&=
\sum_{s\in\mathcal S_T}\ell_{is,0}^* \Big[(\sigma_s-1)\delta_s\,d\ln P_{st} + (\sigma_s-1)\delta_s\,d\ln E_{ist} - \sigma_s\delta_s\,d\ln w_{it}\Big]
\nonumber\\[-0.25em]
&\quad\;\;+\;\ell_{i0}^* \Big[(\sigma_0-1)\delta_0\,d\ln V_{it} - \sigma_0\delta_0\,d\ln w_{it}\Big].
\label{eq:MC-lin}
\end{align}
Collecting the wage terms defines the key denominator 
\begin{equation*}
\lambda_i \;\equiv\; \phi_i \;+\; \sum_{s\in\mathcal S}\ell_{is,0}^*\,\sigma_s\,\delta_s \;>\;0,
\end{equation*}
so that
\begin{equation}
\label{eq:dlnw-solved}
d\ln w_{it}
\;=\;
\frac{1}{\lambda_i}
\left\{
\sum_{s\in\mathcal S_T}\ell_{is,0}^*(\sigma_s-1)\delta_s\,dp_{st}
+
\sum_{s\in\mathcal S}\ell_{is,0}^*(\sigma_s-1)\delta_s\,de_{ist}
-
d\nu_{it}
\right\}.
\end{equation}

\paragraph{Price co-movement.}
Allow other tradable prices to co-move with sector $q$:
\begin{equation}
\label{eq:comove}
dp_{st}\;=\;\rho_{qs}\,dp_{qt}+du_{st},\qquad \rho_{qq}=1,
\end{equation}
where $du_{st}$ is orthogonal to $dp_{qt}$.
Holding $de$ and $d\nu$ orthogonal to $dp_{qt}$, the wage elasticity to the sector-$q$ price is
\begin{equation}
\label{eq:wage-price-elast}
\frac{\partial \ln w_{it}}{\partial \ln P_{qt}}
\;=\;
\frac{1}{\lambda_i}\sum_{s\in\mathcal S_T}\ell_{is,0}^*(\sigma_s-1)\delta_s\,\rho_{qs}.
\end{equation}

\medskip

\begin{proposition}[Heterogeneous first stage
\label{prop:heterogenous_first_stage}]
Under the model described above, sector-$q$ output in region $i$ admits the local levels approximation
\begin{equation}
\label{eq:prop-simple}
X_{iqt}
\;\approx\;
\alpha_{iq}
\;+\;
\kappa_{iq}\bigl[A_{iq}\,p_{qt}\bigr]\
+\;
\varepsilon_{iqt},
\end{equation}
where the heterogeneous coefficient multiplying $A_{iq}$ is
\begin{equation}
\label{eq:bar-kappa}
\kappa_{iq}
\;\equiv\;
\underbrace{\bar E_{iq}\,K_{iq}^{\,1-\theta_q}\,(L_{iq,0})^{\,\theta_q}}_{\text{pre-period capacity (fixed at baseline)}}
\;\times\;
(\sigma_q-1)\,\theta_q\,\delta_q
\left[
1 - \frac{\sigma_q}{\sigma_q-1}\cdot \frac{1}{\lambda_i}
\sum_{s\in\mathcal S_T}\ell_{is,0}^*(\sigma_s-1)\delta_s\,\rho_{qs}
\right].
\end{equation}
with $\lambda_i=\phi_i+\sum_{s\in\mathcal S}\ell_{is,0}^*\sigma_s\delta_s$ and $\rho_{qs}$ from \eqref{eq:comove}. 

\end{proposition}

\medskip

\noindent\textbf{Proof:} See Appendix \ref{appendix_a}.

\begin{remark}[Monotonicity]
\label{remark:monotonicity}
Define the exposure index
\[
S_{iq}\;\equiv\;\frac{1}{\lambda_i}\sum_{s\in\mathcal S_T}\ell_{is,0}^*(\sigma_s-1)\delta_s\,\rho_{qs},
\]
so that the elasticity of sector-$q$ output with respect to its own price is
\[
\frac{\partial \ln X_{iqt}}{\partial \ln P_{qt}}
\;=\;
(\sigma_q-1)\theta_q\,\delta_q\Bigl[\,1-\tfrac{\sigma_q}{\sigma_q-1}\,S_{iq}\Bigr].
\]

Since $A_{iq}>0$ and the capacity term in~\eqref{eq:bar-kappa} is strictly positive, 
monotonicity (a positive first stage) is equivalent to
\[
\kappa_{iq}>0
\quad\Longleftrightarrow\quad
\frac{\partial \ln X_{iqt}}{\partial \ln P_{qt}}>0
\quad\Longleftrightarrow\quad
S_{iq}\;<\;\frac{\sigma_q-1}{\sigma_q}.
\]
If this inequality fails, the general-equilibrium wage effect dominates and the first stage slope
$A_{iq}\kappa_{iq}$ becomes negative.
\end{remark}

Intuition: when prices in other tradable sectors co-move with $p_{qt}$ (large $\rho_{qs}$), those sectors also expand and bid up the local wage. This wage-drag offsets the direct effect of $p_{qt}$ on sector $q$, and can overturn it if the co-moving bundle is strong enough.

The wage-drag term $\tfrac{\sigma_q}{\sigma_q-1}S_{iq}$ is large when:
i) Local labor supply is inelastic ($\phi_i$ small), so $\lambda_i$ is small and the wage moves a lot; ii) the region is highly exposed to co-moving tradables: $\sum_s \ell_{is,0}^*\,\rho_{qs}$ is large (e.g., concentrated mining hubs; many sectors rise together); iii) demand/congestion elasticities are high in those sectors (large $\sigma_s$ or $\delta_s$), amplifying their labor pull; and iv) co-movement is tight and positive ($\rho_{qs}\approx 1$ for many $s$). Conversely, the first stage remains positive when the region is diversified (small $\ell_{is,0}^*$ on co-moving sectors), labor supply is elastic (large $\phi_i\Rightarrow \lambda_i$ large), or co-movement is weak/negative (small or negative $\rho_{qs}$). If one allows nontradables to co-move as well (a local-demand channel), this enters exactly like an additional term in the sum (with $\rho_{q0}>0$) and pushes toward violation.

\begin{remark}[Economic interpretation of $\alpha_{iq}$]
With the normalization $p_{st}=\ln P_{st}-\ln\bar P_s$ and $e_{ist}=\ln E_{ist}-\ln\bar E_{is}$, the linearization in \eqref{eq:prop-simple} is taken around the pre-period baseline $(\bar P,\bar E)$.
The intercept equals the baseline level of sector-$q$ output in region $i$:
\[
\alpha_{iq}
\;=\;
X_{iq,0}
\;=\;
A_{iq}\,\bar E_{iq}\,K_{iq}^{\,1-\theta_q}\,(L_{iq,0})^{\,\theta_q}.
\]
 It is the quantity produced when the centered log price satisfies $p_{qt}=0$ and the centered productivity shock satisfies $e_{iqt}=0$ (i.e., at baseline prices and productivity). Economically, $\alpha_{iq}$ captures time-invariant \emph{capacity} combining geology $A_{iq}$, installed capital $K_{iq}$, and the baseline labor allocation $L_{iq,0}$ determined in the pre-period equilibrium across all sectors. 
\end{remark}

\section{Potential Outcomes Framework and Causal Interpretation}
\label{sec:potential_outcomes}
In this section, we build on the results from the model presented in Section \ref{sec:economic_model}
 and derive the causal interpretation of point estimates from two popular approaches towards the price-exposure design. 

The starting point is to define potential outcomes in the price-exposure setting. The economic model with multiple sectors motivates a potential outcomes model which depends on multiple shifters. Let $Y_{it}(x_{1},...,x_{S})$ denote the potential outcome of region $i$ in period $t$ that would occur if the outputs of the $S$ sectors were set to $(x_{1},...,x_{S})$. We assume that potential outcomes are linear functionals of the outputs in each sector:

\begin{equation}
\label{eq:potential_outcomes}
  Y_{it}(x_{1},...,x_{S})=\sum_{s=1}^{S}x_{s}\beta_{is}+\eta_{it}
\end{equation}

Under this model, we have $\eta_{it}=Y_{it}(0,...,0)$ as the potential outcome associated to output 0 across all sectors and $\beta_{iq}$ as the causal effect of an increase of the output from sector $q$ by one unit: 
\begin{equation*}
  \frac{\partial Y_{it}(x_{1},...,x_{S})}{\partial x_{q}}=\beta_{iq}
\end{equation*}

We follow AKM and take a finite-population approach towards identification and inference. That is, we define the population of interest to be the observed set of $N$ regions over $T$ periods of time, instead of focusing on a large superpopulation over from which the observations are drawn. Uncertainty in the setting arises from the stochastic nature of the shock, while causal effects and potential outcomes are taken as given. Define $\mathcal{F}_0 \equiv \{ \eta_{it},\varepsilon_{ist},\beta_{is},\alpha_{is},\kappa_{is} \}_{i=1,t=1,s=1}^{N,T,S}$.
 as the collection of parameters that determine the panel of causal effects and potential outcomes. We study the property of different estimators conditional on $\mathcal{F}_{0}$, under the assumption that the price of the sector of interest is exogenous. Our setup can be summarized by the two following assumptions:

\begin{assumption}[Potential outcomes and first stage]
\label{assumption:potential_outcomes}
\mbox{}\\
(i) The observed outcomes are generated by
\[
Y_{it} \;=\; Y_{it}(x_{1},..., x_{q},...,x_{S}),
\]
where $Y_{it}(\cdot)$ follows the specification in equation~\eqref{eq:potential_outcomes}.\\
(ii) Output in each sector is determined according to the structure in Section~\ref{sec:economic_model}, so that Proposition~\ref{prop:heterogenous_first_stage} holds.
\end{assumption}

Assumption~\ref{assumption:potential_outcomes} can be seen as the usual exclusion restriction in this setting. Part (i) says that prices do not have a direct effect on $Y_{it}$. Instead, they only affect the outcome to the extent that they affect outputs. In particular, we rule out direct regional effects of $p_{qt}$ on the outcome even holding the sectoral outputs fixed, such as direct local political responses to commodity price shocks. This corresponds to the common applied interpretation that the price-exposure instrument identifies the effect of ``sector $q$ activity`` as long as there are no other direct channels of the price on the outcome. Thus Assumption~\ref{assumption:potential_outcomes}(i) is purely a restriction on the way outcomes depend on sectoral activity; it does not by itself impose any statistical exogeneity condition on the price process.\footnote{Applied work often invokes small-open-economy intuition or external instruments for the commodity price to argue that the country is a price taker and local conditions do not feed back into $p_{qt}$. That kind of argument speaks to a separate exogeneity condition (in Assumption~\ref{assumption:random_price} below), not to the exclusion restriction as stated in Assumption~\ref{assumption:potential_outcomes}(i).}

 What Assumption~\ref{assumption:potential_outcomes}(i) does not restrict is how a shock to $p_{qt}$ propagates across sectors: our multi-sector model implies that a change in $p_{qt}$ will generally move outputs in many sectors, through both general-equilibrium labor reallocation and co-movement in other tradable prices. Whenever $Y_{it}$ depends on these additional sectors (i.e., $\beta_{is}\neq 0$ for some $s\neq q$), standard price-exposure regressions implicitly combine the effect of sector-$q$ activity with effects operating through other sectors. Characterizing how this multi-sector structure contaminates the usual estimands is one of our goals in this paper.

At the same time, in principle Assumption~\ref{assumption:potential_outcomes}(i) need not require that there is a single economic mechanism from sector $q$ to $Y_{it}$. In practice, the outcome might be affected by several economic channels related to sector $q$ (such as employment or income in the gold mining sector), and not only by the actual output $X_{iqt}$ (the amount of gold produced). The outcome is allowed to depend on employment, income, or other variables in sector $q$, as long as these are (possibly complicated) functions of $X_{iqt}$; formally, these channels are absorbed into the response function $Y_{it}(\cdot)$. The restrictive component of the assumption, which we adopt for simplicity, is the linear form in equation~\eqref{eq:potential_outcomes}, which compresses all such mechanisms for each sector into a one-dimensional sufficient statistic $X_{iqt}$ and an additive effect $\beta_{iq}$. This is usually acknowledged by applied researchers studying the effect of ``sector-$q$ activity'', but it underscores the need to think carefully about how different choices of proxies for $X_{iqt}$ (e.g., mined area, physical output, or sectoral employment) map into the economic effect that the resulting estimates are taken to represent.

\begin{assumption}[Price exogeneity]
\label{assumption:random_price}
  The price in sector $q$ is as good as randomly assigned, in the sense that for all $t\in\left\{1,...,T\right\}$,

  \begin{equation*}
    \mathbb{E}\left[p_{qt}|\mathcal{F}_{0}\right]=0
  \end{equation*}
\end{assumption}

Assumption \ref{assumption:random_price} imposes that the price shock in sector $q$ is mean-independent from potential outcomes, given the mean-independence from $\eta_{it}$ and the vector $(\beta_{i1},...,\beta_{iS})$, and also mean-independence from the potential ``treatments'' in each sector, which follows from mean-independence from $\{ (\alpha_{is},\kappa_{is},\varepsilon_{ist}) \}_{i,s,t}$.
 Thus, Assumption \ref{assumption:random_price} can be interpreted as the standard exogeneity assumption in IV settings that the instrument is mean-independent from potential quantities, but expressed in terms of the parameters from the stylized economic model.\footnote{This assumption is substantively strong. Our multi-sector model suggests that general-equilibrium labor reallocation will cause outputs in other sectors to respond to changes in $p_{qt}$, even when the other sectoral prices $\{p_{st}\}_{s\neq q}$ are statistically independent of $p_{qt}$. Requiring mean-independence from the entire vector $\{\varepsilon_{ist}\}_{i,s,t}$ effectively ignores this channel. It amounts to imposing $\mathbb{E}\!\left[p_{qt}\varepsilon_{ist}\mid\mathcal{F}_{0}\right]=0$ for all $i,s,t$, so any component of $X_{ist}$ that moves systematically with $p_{qt}$ must be captured in the structured part of the first stage rather than in $\varepsilon_{ist}$. In other words, we allow the outcome to move with the output in sector $s \neq q$, but only through co-movement of $p_{st}$ with $p_{qt}$, ignoring the channel of output changes via changes in wages. We derive our main results under this assumption for expositional simplicity. In Section~\ref{sec:endogenous_epsilon} we show how allowing such general-equilibrium terms to enter $\varepsilon_{ist}$ generates additional contamination even when sectoral prices are mutually independent.} In applications, this is typically motivated by viewing the country as a small open economy that takes $p_{qt}$ as given, or by constructing an external instrument for the commodity price (such as with shocks to global demand or to supply in other countries).

% interpretacao boa pra essa footnote sobre o epsilon, que parece estar impondo uma exclusion restriction a mais: You’re still allowing Y to depend on coffee; you’re just saying the only way Pq moves coffee is via coffee’s own price comoving with gold, not via wages.

\subsection{IV Estimands}

We begin with the causal interpretation of IV estimates, which exploit the price-exposure as an instrument for output in sector $q$. We define the instrument as $Z_{iqt}\equiv A_{iq}\times p_{qt}$, where $p_{qt} \equiv \ln P_{qt} - \ln\bar P_{qt}$.\footnote{Results are equivalent without demeaning the price, as long as time fixed effects are included.} In this setting, the 2SLS estimator stacked across time-periods of the coefficient on $X_{iqt}$ is given by 

\begin{equation*}
  \widehat{\beta}_{q}^{2SLS}=\frac{\sum_{i=1}^{N}\sum_{t=1}^{T}Z_{iqt}Y_{it}}{\sum_{i=1}^{N}\sum_{t=1}^{T}Z_{iqt}X_{iqt}}
\end{equation*}

Given the finite-population framework, the 2SLS estimand is defined as the population analogue of the 2SLS estimator under repeated sampling of the shocks:

\begin{equation*}
  \beta_{q}^{2SLS}=\frac{\sum_{i=1}^{N}\sum_{t=1}^{T}\mathbb{E}\left [ Z_{iqt}Y_{it}|\mathcal{F}_{0} \right ]}{\sum_{i=1}^{N}\sum_{t=1}^{T}\mathbb{E}\left [ Z_{iqt}X_{iqt}|\mathcal{F}_{0} \right ]}
\end{equation*}

The following proposition establishes the causal interpretation of the 2SLS estimand:

\begin{proposition}
\label{prop:beta_2sls}
  Suppose that Assumption \ref{assumption:random_price} holds and the potential outcomes model is correctly specified (i.e. Assumption \ref{assumption:potential_outcomes} holds). Then,

  \begin{equation*}
    \beta_{q}^{2SLS}=\frac{\sum_{i=1}^{N}\sum_{t=1}^{T}\pi_{iqt}\beta_{iq}}{\sum_{i=1}^{N}\sum_{t=1}^{T}\pi_{iqt}}+\frac{\sum_{i=1}^{N}\sum_{t=1}^{T}\left\{ \sum_{s\neq q}\pi_{isqt}\beta_{is}\right\}}{\sum_{i=1}^{N}\sum_{t=1}^{T}\pi_{iqt}}
  \end{equation*}
\\
  where $\pi_{iqt}=\kappa_{iq}A_{iq}^{2}\mathbb{V}\left ( p_{qt}|\mathcal{F}_{0} \right )$ and $\pi_{isqt}=\kappa_{is}A_{is}A_{iq}\mathbb{C}ov\left ( p_{qt},p_{st}|\mathcal{F}_{0} \right )$.
\end{proposition}

\noindent\textbf{Proof:} See Appendix \ref{appendix_b}.

Proposition \ref{prop:beta_2sls} decomposes the IV estimand in two terms. The first is a weighted average of region-specific parameters associated to an increase in the output of sector $q$. The weights are increasing in the region-specific elasticity $\kappa_{iq}$ and the region-specific exposure $A_{iq}$. The weights are positive as long as $\kappa_{iq}\geq 0$, which can be interpreted as the monotonicity condition from IV settings mapped to the economic model from Section \ref{sec:economic_model} (see Remark \ref{remark:monotonicity}).

The second term is a linear combination of region-specific parameters associated to increases in the output of all the other sectors stemming from movements in other prices. The sign of the coefficients associated to these causal parameters depends on the region-specific elasticity of the sectors, $\left\{\kappa_{is}\right\}_{s\neq q}$ and the covariances between the price of sector $q$ and prices of other sectors. The 2SLS estimand can only be interpreted as a weighted average of causal effects in the case where the prices of all sectors are independent from the price in sector $q$. If all elasticities and price covariances are positive --- and thus the monotonicity condition in Remark \ref{remark:monotonicity} holds ---, then the 2SLS estimand holds a weakly causal interpretation in the \citet{blandhol2022tsls} sense.

In the case of panel data it is common to include time fixed effects to account for changes in the mean of the shift over time. Adding time fixed effects is equivalent to demeaning the price shock. Also, note that serial correlation in international prices does not affect the causal interpretation of the estimand, as the causal model rules out dynamic effects.

\subsubsection{Causal Interpretation under General Equilibrium Effects}
\label{sec:endogenous_epsilon}

Proposition \ref{prop:beta_2sls} provides the causal interpretation of the 2SLS estimand under the light of the economic model from Section \ref{sec:economic_model}. It follows from the causal decomposition that the contamination from causal effects of the output from other sectors affects the interpretation of the 2SLS estimand in the presence of price co-movement. The model, however, rules general equilibrium effects outside of price co-movement which might be relevant in applied settings.

In this section, we derive the causal interpretation of the 2SLS estimand considering a general equilibrium channel in which the prices of sector $q$ affect the output in other sectors. To that end, we write the variable $\varepsilon_{ist}$ as $\varepsilon_{ist}=\tilde{\varepsilon}_{ist}+\sum_{s'\neq s}\gamma_{iss'}p_{s't}$, where $\tilde{\varepsilon}_{ist}$ is idiosyncratic and the vector of parameters $\gamma_{iss'}$ captures how the output in sector $s$ at time $t$ responds to changes in prices from other sectors at the same time period.

Under these new specification of the model, we define the filtration as

\begin{equation*}
  \tilde{\mathcal{F}}_{0}=\left\{\eta_{it},\alpha_{is},\beta_{is},\left ( \gamma_{iss'} \right )^{'},\kappa_{is},A_{is},\tilde{\varepsilon}_{ist} \right\}_{i=1,t=1,s=1}^{N,T,S}
\end{equation*}

The exogeneity assumption we invoke is $\mathbb{E}\left[p_{st}|\tilde{\mathcal{F}}_{0}\right]=0$. In the next proposition, we show that even when prices are independent, the 2SLS estimand can be decomposed as a weighted average of causal effects associated to the output in sector $q$ and a contamination term.

\begin{proposition}
\label{prop:beta_2sls_general_eq}
  Suppose Assumption \ref{assumption:potential_outcomes} holds and that $\mathbb{E}\left[p_{st}|\tilde{\mathcal{F}}_{0}\right]=0$ for all $s=1,...,S$. Then,

\begin{equation*}
  \beta_{q}^{2SLS}=\frac{\sum_{i=1}^{N}\sum_{t=1}^{T}\tilde{\pi}_{iqt}\beta_{iq}}{\sum_{i=1}^{N}\sum_{t=1}^{T}\tilde{\pi}_{iqt}}+\frac{\sum_{i=1}^{N}\sum_{t=1}^{T}\left\{ \sum_{s\neq q}\tilde{\pi}_{isqt}\beta_{is}\right\}}{\sum_{i=1}^{N}\sum_{t=1}^{T}\tilde{\pi}_{iqt}}+\frac{\sum_{i=1}^{N}\sum_{t=1}^{T}\left\{ \sum_{s\neq q}A_{iq}\gamma_{isq}\mathbb{V}\left(p_{qt}|\tilde{\mathcal{F}}_{0}\right)\beta_{is}\right\}}{\sum_{i=1}^{N}\sum_{t=1}^{T}\tilde{\pi}_{iqt}}
\end{equation*}

  where $\tilde{\pi}_{iqt}=\kappa_{iq}A_{iq}^{2}\mathbb{V}\left ( p_{qt}|\tilde{\mathcal{F}}_{0} \right )$ and $\tilde{\pi}_{isqt}=\kappa_{is}A_{is}A_{iq}\mathbb{C}ov\left ( p_{qt},p_{st}|\tilde{\mathcal{F}}_{0} \right )$.

\end{proposition}

\noindent\textbf{Proof:} See Appendix \ref{appendix_b}.

Proposition \ref{prop:beta_2sls_general_eq} shows that if the prices of sector $q$ directly affect the output in other sectors, then the 2SLS estimand does not hold a straightforward causal interpretation even in the absence of price co-movement. The estimand holds a weakly causal interpretation if for all $s\neq q$, the parameter $\gamma_{isq}$ is nonnegative. That is, if an increase in the price of sector $q$ affects positively the output in all sectors (or at least does not affect any sector negatively), then the estimand holds a weakly causal interpretation.

\subsubsection{Sensitivity Analysis}

Propositions \ref{prop:beta_2sls} and \ref{prop:beta_2sls_general_eq} imply that the exogeneity of the price of sector $q$ is not sufficient to provide a clear causal interpretation to the IV estimand. The two possible sources of contamination are (i) causal effects from other sectors affecting the estimand, which is driven by price co-movement and (ii) the effect of the price in sector $q$ affecting the output of other sectors.

In the case where the prices and outputs from other sectors are observed by the applied researcher, the natural solution is to control for such quantities and close the channels of potential contamination of the estimand. Nevertheless, it is often the case in price-exposure designs that these quantities are not available for the researcher. Many applications, especially in the development economics literature, focus on the interaction between informal or illegal markets and legal markets, and is often the case that prices and outputs in illegal markets are not available.

Our framework motivates a simple procedure for sensitivity analysis based on the 2SLS estimand. When the sources of contamination cannot be observed in the data, researchers may place assumptions on the magnitude of the bias and obtain identified sets for the weighted average of causal effects from sector $q$ and conduct inference on the partially identified causal estimand $\beta_{q}$.

Concretely, assume the contamination terms lie in the interval $\left[\underline{b}, \overline{b}\right]$\footnote{The interpretation of these sensitivity parameters is straightforward in the case where only one of the channels of contamination poses a threat to the causal interpretation of the estimand. Nevertheless, the procedure is valid in all cases.}. Propositions \ref{prop:beta_2sls} and \ref{prop:beta_2sls_general_eq} imply that $\beta_{q}$ lies in the interval $\left[\beta_{q}^{LB},\beta_{q}^{UB}\right]$, where

\begin{equation*}
  \beta_{q}^{LB}=\beta_{q}^{2SLS}-\overline{b},\ \beta_{q}^{UB}=\beta_{q}^{2SLS}-\underline{b}
\end{equation*}

A natural estimation procedure is to plug-in the 2SLS estimator $\widehat{\beta}_{q}^{2SLS}$ in the expression above, which yields unbiased estimates for the bounds. By choosing different values for $\underline{b}$ and $\overline{b}$, researchers can assess the robustness of the qualitative takeaways of the study.

\subsection{Two-Way Fixed Effects Estimands}

Another pervasive approach in the applied literature is to specify a two-way fixed effect (TWFE) regression, with time ($\theta_{t}$) and individual ($\gamma_{i}$) fixed effects, and the price-exposure variable:

\begin{equation*}
  Y_{it}=\theta_{t}+\gamma_{i}+\beta^{TWFE}\left(A_{iq}p_{qt}\right)+u_{it}
\end{equation*}

For the sake of simplicity, we consider the case of panel data with two periods of time, so that the regression model can be expressed as

\begin{equation*}
  \Delta Y_{i}=(\theta_{2}-\theta_{1})+\beta^{TWFE}A_{iq}\Delta p_{q}+\Delta u_{i}
\end{equation*}

and the TWFE coefficient associated to the price-exposure variable as

\begin{equation*}
  \widehat{\beta}_{q}^{TWFE}=\frac{\sum_{i=1}^{N}\Delta Z_{iq}\Delta Y_{i}}{\sum_{i=1}^{N}\left ( \Delta Z_{iq} \right )^{2}}
\end{equation*}

The TWFE estimand is defined as the population analogue of the TWFE estimator under repeated sampling of the shocks:

\begin{equation*}
  \beta^{TWFE}=\frac{\sum_{i=1}^{N}\mathbb{E}\left [ \Delta Z_{iq}\Delta Y_{i}|\mathcal{F}_{0} \right ]}{\sum_{i=1}^{N}\mathbb{E}\left [ \left(\Delta Z_{iq}\right)^{2}|\mathcal{F}_{0} \right ]}
\end{equation*}

It is common for applied researchers to claim a 'reduced form' or Intention-to-Treat (ITT) interpretation for the TWFE estimand. The proposition below provides the causal interpretation of the estimand:

\begin{proposition}
\label{prop:beta_twfe}
  Suppose that Assumption \ref{assumption:random_price} holds and the potential outcomes model is correctly specified. Then, 

  \begin{equation*}
    \beta^{TWFE}=\frac{\sum_{i=1}^{N}\kappa_{iq}(A_{iq})^{2}\mathbb{V}(\Delta p_{q}|\mathcal{F}_{0})\beta_{iq}}{\sum_{i=1}^{N}(A_{iq})^{2}\mathbb{V}(\Delta p_{q}|\mathcal{F}_{0})}+\frac{\sum_{i=1}^{N}\left\{ \sum_{s\neq q}\kappa_{is}A_{is}A_{iq}\mathbb{C}ov(\Delta p_{s},\Delta p_{q}|\mathcal{F}_{0})\beta_{is}\right\}}{\sum_{i=1}^{N}(A_{iq})^{2}\mathbb{V}(\Delta p_{q}|\mathcal{F}_{0})}
  \end{equation*}
\end{proposition}

\noindent\textbf{Proof:} See Appendix \ref{appendix_b}.

Proposition \ref{prop:beta_twfe} shows that the TWFE can be decomposed as a linear combination of causal parameters. The first term is a linear combination of region-specific effects of an increase in the output of sector $q$. The linear combination holds a weakly causal interpretation as long as monotonicity in the first-stage holds. However, it can be interpreted as a convex weighted average of causal effects only in the case when the $A_{iq}^{2}$-weighted average of $\kappa_{iq}$ equals one
(which holds, for example, if $\kappa_{iq}=1$ for all regions). The second term is a contamination term similar to the one in Propostion \ref{prop:beta_2sls}, but depends on the covariance of the variation of prices across time. 

Hence, it follows that the TWFE estimand holds the interpretation of a weighted average of causal effects only under additional restrictions on the first-stage (which is why researchers usually claim a reduced form interpretation towards this estimand) and an additional assumption regarding the covariance of prices from different sectors.

Nevertheless, it is worth noting that the causal decomposition from Proposition \ref{prop:beta_twfe} can be obtained under an exogeneity assumption weaker than Assumption \ref{assumption:random_price}. By exploiting within-regions variation across time, the time-invariant parameters from the model such as $\left\{\alpha_{is}\right\}_{s=1}^{S}$ are cancelled out. Moreover, define $\Delta \mathcal{F}_{0}\equiv\left\{ \Delta\eta_{i},\Delta\varepsilon_{is},\beta_{is},\kappa_{is} \right\}_{i=1,s=1}^{N,S}$ to invoke the following exogeneity assumption:

\begin{assumption}[Price exogeneity in trends]
\label{assumption:random_delta_price}
  The price in sector $q$ is randomly assigned in the sense that 

  \begin{equation*}
    \mathbb{E}\left[\Delta p_{q}|\Delta \mathcal{F}_{0}\right]=0
  \end{equation*}
\end{assumption}

Assumption \ref{assumption:random_delta_price} is weaker than Assumption \ref{assumption:random_price}, and can be interpreted as a parallel trends assumption for potential outcomes and potential output. It is different than the usual parallel trends assumption, however, as it assumes exogeneity conditional on the parameters $\left\{\beta_{is}\right\}_{s=1}^{S}$ and $\left\{\kappa_{is}\right\}_{s=1}^{S}$, which usually is not the case in TWFE/Two-stage TWFE settings. This is necessary because unlike the standard DiD settings, there is no time-period in which regions are fully untreated nor a time-period where the instrument has value zero for all regions. Contrary to a common interpretation in empirical work, the reduced form price-exposure design allows for non-parallel trends in \(Y_{it}\) across regions with different exposure levels, and the exposures \(A_i\) may be endogenous as long as innovations in \(p_{qt}\) are mean-independent of the finite-population primitives. Below, we show that the estimand $\beta^{TWFE}$ maintains its causal decomposition under the weaker exogeneity assumption.

\begin{corollary}
\label{corollary:beta_twfe}
  Suppose Assumption \ref{assumption:random_delta_price} holds. Then,

  \begin{equation*}
    \beta^{TWFE}=\frac{\sum_{i=1}^{N}\kappa_{iq}(A_{iq})^{2}\mathbb{V}(\Delta p_{q}|\Delta\mathcal{F}_{0})\beta_{iq}}{\sum_{i=1}^{N}(A_{iq})^{2}\mathbb{V}(\Delta p_{q}|\Delta\mathcal{F}_{0})}+\frac{\sum_{i=1}^{N}\left\{ \sum_{s\neq q}\kappa_{is}A_{is}A_{iq}\mathbb{C}ov(\Delta p_{s},\Delta p_{q}|\Delta\mathcal{F}_{0})\beta_{is}\right\}}{\sum_{i=1}^{N}(A_{iq})^{2}\mathbb{V}(\Delta p_{q}|\Delta\mathcal{F}_{0})}
  \end{equation*}
\end{corollary}

Corollary \ref{corollary:beta_twfe} provides a straightforward motivation for the TWFE approach towards the price-exposure design. Despite the fact that in general it only holds a weakly causal interpretation, the estimand holds its causal interpretation under a weaker exogeneity assumption than the one necessary for the IV estimand. In general, researchers choose the TWFE approach when they have access to the price-exposure variable but not to the actual variable of interest, which is why it is often referred to as a reduced form approach. Proposition \ref{prop:beta_twfe} and Corollary \ref{corollary:beta_twfe} formalize such an interpretation in terms of the stylized economic model from Section \ref{sec:economic_model}.

\section{Estimation and Inference}
\label{sec:estimation_inference}
In this section, we analyze the statistical properties of the 2SLS estimator $\widehat{\beta}_{q}^{2SLS}$. We consider large-sample properties of $\widehat{\beta}_{q}^{2SLS}$ as the number of regions grows to infinity ($N\rightarrow\infty$), and the number of time periods grows to infinity ($T\rightarrow\infty$). Here we describe the main assumptions, technical regularity conditions are stated in Appendix \ref{appendix_c}. Alongside Assumptions \ref{assumption:potential_outcomes} and \ref{assumption:random_price}, we invoke the following assumption regarding the panel of sector prices:

\begin{assumption}[Price independence]
\label{assumption:independent_prices}
\mbox{}\\
  (i) For all $t\in\left\{1,...,T\right\}$, the prices $\left(p_{1t},...,p_{St}\right)$ are independent conditional on $\mathcal{F}_{0}$; \\ (ii) For all $s\in\left\{1,...,S\right\}$, the prices $\left(p_{s1},...,p_{sT}\right)$ are independent conditional on $\mathcal{F}_{0}$.
\end{assumption}

Assumption \ref{assumption:independent_prices} (i) requires sector prices to be independent. Under this assumption, it follows that $\mathbb{C}ov(p_{qt},p_{st}|\mathcal{F}_{0})=0$ for all $s$, and thus the 2SLS estimand captures the weighted average of sector $q$ causal effects without the contamination terms that affect its causal interpretation. Assumption \ref{assumption:independent_prices} (ii) requires the sector prices to be independent across time periods. These assumptions combined extend to our setting the assumption underlying randomization style inference in randomized controlled trials that the treatment assignment is independent across units.

We now establish the asymptotic properties of the 2SLS is consistent and asymptotically normal, with $\sqrt{NT}$ convergence rate.

\begin{theorem}
\label{theorem:asymptotic_variance}
  Suppose Assumptions \ref{assumption:potential_outcomes}, \ref{assumption:random_price}, \ref{assumption:independent_prices}, and Assumptions \ref{assumption:regularity_consistency} and \ref{assumption:regularity_normality} from Appendix \ref{appendix_c} hold. Then,

  \begin{equation*}
    \widehat{\beta}_{q}^{2SLS}=\beta_{q}^{2SLS}+o_{p}(1)
  \end{equation*}

  and

  \begin{equation*}
  \sqrt{NT}\left ( \widehat{\beta}_{q}-\beta_{q} \right )\overset{d}{\rightarrow}\mathcal{N}\left ( 0,\frac{\mathcal{V}_{NT}}{\left ( \frac{1}{NT}\sum_{i=1}^{N}\sum_{t=1}^{T}Z_{iqt}X_{iqt} \right )^{2}} \right )
\end{equation*}

where

\begin{equation*}
  \mathcal{V}_{NT}=\frac{1}{NT}\mathbb{V}\left ( \sum_{i=1}^{N}\sum_{t=1}^{T}(A_{iq}p_{qt})u_{it}|\mathcal{F}_{0} \right )
\end{equation*}

with $u_{it}$ being the residual in the structural equation.
\end{theorem}

\noindent\textbf{Proof:} See Appendix \ref{appendix_c}.

Theorem \ref{theorem:asymptotic_variance} shows that the asymptotic variance has the usual 'sandwich' format. To construct a consistent estimator for the standard error, it is sufficient to construct a consistent estimator for $\mathcal{V}_{NT}$. Consider first the case of homogeneous treatment effects, $\beta_{iq}=\beta$. Then it follows under Assumptions \ref{assumption:random_price} and \ref{assumption:independent_prices} that

\begin{equation*}
  \mathcal{V}_{NT}=\frac{1}{NT}\sum_{t=1}^{T}\mathbb{V}\left ( p_{qt}|\mathcal{F}_{0} \right )\left ( \sum_{i=1}^{N}A_{iq}u_{it} \right )^{2}
\end{equation*}

which motivates the following plug-in estimator for the variance:

\begin{equation*}
  \widehat{V}_{PE}(\widehat{\beta})=\frac{\widehat{\mathcal{V}}_{NT}}{\left ( \frac{1}{NT}\sum_{i=1}^{N}\sum_{t=1}^{T}Z_{iqt}X_{iqt} \right )^{2}}
\end{equation*}

where $\widehat{\mathcal{V}}_{NT}=\frac{1}{NT}\sum_{t=1}^{T}p_{qt}^{2}\left ( \sum_{i=1}^{N}A_{iq}\widehat{u}_{it} \right )^{2}$. The estimator for the variance is similar to the one proposed by AKM, but instead of exploring variation across sectors, we exploit variation across time periods. The variance estimate leads to valid inference in the case of constant treatment effects, and inference remains valid under heterogeneous effects under additional regularity conditions stated in Appendix \ref{appendix_c}.

As in AKM, our asymptotic results point to a possible overrejection problem associated to standard inference procedures. Take, for instance, the Eicker-Huber-White (or cluster-robust) standard errors. The expected difference between our proposed variance estimator and the robust variance estimator is a function of the correlation between residuals across regions and the exposures of these regions. Thus, if residuals are positively correlated, then it leads to an overrejection problem for the EHW variance estimator.

As AKM point out, "standard inference methods may overreject even when the unobserved shifters (prices from other sectors in our settings) contained in the residual vary along a different dimension than the shift-share (price-exposure in our setting) covariate of
interest" if the variance of prices from other sectors is large and the correlation structure from the exposures in the residual is similar to the structure of the exposure in sector $q$. That seems to be the main channel driving overrejection in our setting. In the next section, we conduct a series of Monte Carlo simulations to assess how large $N$ and $T$ need to be for these asymptotics to provide a good approximation to the finite sample distribution of the 2SLS estimator.

\subsection{Inference for Sensitivity Analysis}

The results from Theorem \ref{theorem:asymptotic_variance} can be combined with existing results from the partial identification literature to produce valid confidence intervals for the partially identified causal parameter $\beta_{q}$. We adapt the confidence intervals from \citet{imbensmanski} for our setting. Let $\mathcal{B}=\overline{b}-\underline{b}$ denote the length of the identified set. The Imbens-Manski confidence interval in the price-exposure setting takes the form

\begin{equation*}
  \left[ \widehat{\beta}^{2SLS}_{q}-C\sqrt{\widehat{V}_{PE}(\widehat{\beta})},\widehat{\beta}^{2SLS}_{q}+C\sqrt{\widehat{V}_{PE}(\widehat{\beta})} \right]
\end{equation*}

where $C$ is a constant that solves $\Phi\left( \frac{\mathcal{B}}{\sqrt{\widehat{V}_{PE}}}+C \right)-\Phi\left( -C \right)=1-\alpha$, with $\Phi(.)$ being the standard normal CDF. The Imbens-Manski confidence interval can be used to conduct breakdown analysis in order to assess the robustness of qualitative takeaways from the study.

The breakdown point $b^{*}$ is the value of contamination required to overturn a causal conclusion. For instance, how large must the contamination be in order for the Imbens-Manski
interval to contain a null effect.

Standard error estimates for $\widehat{\beta}_{q}^{2SLS}$ are weakly conservative, and thus Imbens-Manski CIs for the identified set will be conservative as well as estimated breakdown points. See Appendix D of \citet{asheshjon} for a thorough discussion on inference in identified sets under design-based uncertainty.

\section{Monte Carlo Simulations}
\label{sec:montecarlo}
In this section, we conduct a series of Monte Carlo experiments in order to study the finite sample properties of the 2SLS estimator and different inference procedures in price-exposure settings. The main objective of this exercise are (i) to verify that under Assumptions \ref{assumption:potential_outcomes}, \ref{assumption:random_price}, and \ref{assumption:independent_prices} the 2SLS yields an unbiased estimate for the estimand $\beta_{q}$ and (ii) to assess how large the panel must be in order for the asymptotic results from Theorem \ref{theorem:asymptotic_variance} to provide a good approximation for the finite sample distribution of the 2SLS estimator.

In order to do so, we consider the performance of the 2SLS estimator in terms of its mean bias, and the performance of the proposed standard error estimator (we label it \textit{P-E}) and the Eicker-Huber-White estimator (we label it \textit{Robust}) in terms of its empirical 95\% coverage probability and mean standard error size under different sizes of the panel sample. 

We consider a simple data generating process (DGP) with only two tradable sectors. For each of the time periods, the prices from each sector are drawn from uniform distributions with mean zero independent across the sectors and time periods, ensuring that Assumptions \ref{assumption:potential_outcomes}, \ref{assumption:random_price}, and \ref{assumption:independent_prices} hold.

The idiosyncratic shock in the ouptut of each sector, $\varepsilon_{ist}$, follows a Gaussian distribution. The exposures $A_{is}$ are drawn from uniform distributions in which the lower limit is greater than zero. The terms $\alpha_{is}$ are generated by multiplying $A_{is}$ with a positive constant. Finally, the elasticity terms $\kappa_{is}$ are drawn from a standard uniform distribution, which ensures that monotonicity holds.

The untreated potential outcomes $\eta_{it}$ are drawn from a Gaussian distribution. The causal effects $\beta_{is}$ follow a uniform distribution between 0 and 2. Under this definition for the values that comprise $\mathcal{F}_{0}$, the estimand $\beta^{2SLS}_{q}$ is equal to 1.

We document the performance of the standard error estimators over the randomization distribution. We first generate the filtration which generates potential outcomes and then simulate the data through different simulations of price sectors, holding potential treatments fixed. We conduct 1.000 Monte Carlo simulations for each exercise.

The finite sample properties of the 2SLS estimator and the inference procedures are compared under two scenarios. In the first, the number of regions is fixed at $N=1.000$ and we study the performance of the estimators under different numbers of time periods. The results are summarized in Table \ref{tab:montecarlo1}.

In terms of bias, the 2SLS estimator exhibits desirable finite sample properties across all panel specifications. Naturally, the average bias exhibits its small value in magnitude when $T=100$. Yet, even in the worst case scenario with a single time period available, the mean bias is smaller than 1\% of the true causal parameter. In terms of inference, the simulation results illustrate the overrejection problem associated to standard inference procedures. The robust standard errors are substantially smaller than our proposed standard error estimator across all panel sizes. The empirical coverage of the 95\% confidence interval goes from 0 in the cross-sectional case to 62\% in the largest panel. The proposed \textit{P-E} estimator outperforms the robust standard error under in all panels, but only exhibits empirical coverage close to 95\% when the number of time periods reaches triple digits.

\begin{table}[t]
\centering \small
\caption{Monte Carlo Simulations - Scenario 1}
\label{tab:montecarlo1}
\begin{tabular}{cccccccc}
\hline
   & Mean Bias & & \multicolumn{2}{c}{Mean SE} & & \multicolumn{2}{c}{Coverage} \\ \cline{2-2} \cline{4-5} \cline{7-8} 
   & & & Robust    & P-E     & & Robust    & P-E     \\ \hline
$N=1.000, T=1$  & 0.008   & & 0.0     & 0.002    & & 0  & 0 \\
$N=1.000, T=5$  & -0.009  & & 0.048     & 0.519    & & 0.369     & 0.560    \\
$N=1.000, T=10$ & 0.007   & & 0.032     & 0.376    & & 0.463     & 0.717    \\
$N=1.000, T=20$ & 0.001   & & 0.051     & 0.382    & & 0.471     & 0.818    \\
$N=1.000, T=100$ & -0.002  & & 0.015     & 0.310    & & 0.617     & 0.938    \\ \hline
\end{tabular}
 \begin{minipage}{0.9\textwidth}
  \footnotesize
  \medskip
  \footnotesize \noindent \textit{Note:} Simulations based on 1.000 Monte Carlo experiments with sample size $N=1.000$ under the DGP described in the Section."Mean bias", "Mean SE" and "Coverage" stand for the mean simulated bias, mean simulated size of standard errors and 95\% coverage 
 \end{minipage}
\end{table}

In the second scenario, the number of time-periods is fixed at $T=1.000$ and we study the performance of the standard error estimators under different numbers of regions. The results are displayed in Table \ref{tab:montecarlo2}.

\begin{table}[t]
\centering \small
\caption{Monte Carlo Simulations - Scenario 2}
\label{tab:montecarlo2}
\begin{tabular}{cccccccc}
\hline
   & Mean Bias & & \multicolumn{2}{c}{Mean SE} & & \multicolumn{2}{c}{Coverage} \\ \cline{2-2} \cline{4-5} \cline{7-8} 
   & & & Robust    & P-E     & & Robust    & P-E     \\ \hline
$N=1, T=1.000$  & -0.027  & & 0.004     & 0.006    & & 0     & 0 \\
$N=5, T=1.000$  & -0.001  & & 0.023     & 0.049    & & 0     & 0    \\
$N=10, T=1.000$ & 0.010   & & 0.042     & 0.099    & & 0.214     & 0.554    \\
$N=20, T=1.000$ & -0.001  & & 0.031     & 0.126    & & 0.227     & 0.591    \\
$N=100, T=1.000$ & 0.001   & & 0.015     & 0.115    & & 0.263     & 0.961    \\ \hline
\end{tabular}
 \begin{minipage}{0.9\textwidth}
  \footnotesize
  \medskip
  \footnotesize \noindent \textit{Note:} Simulations based on 1.000 Monte Carlo experiments with sample size $N=1.000$ under the DGP described in the Section."Mean bias", "Mean SE" and "Coverage" stand for the mean simulated bias, mean simulated size of standard errors and 95\% coverage 
 \end{minipage}
\end{table}
  
Once again, the average bias of the 2SLS estimator is negligible across the different sizes of the panel. When it comes to inference, the robust standard error estimators yield severely biased confidence intervals across all specifications. The overrejection problem appears to be more dramatic than in the case of a large cross-section and a growing number of time periods. In the best case, the empirical coverage of the 95\% confidence interval is 26\%, almost half of what is observed in the best case under Scenario 1. The simulation results show that the \textit{P-E} standard error estimator only exhibits desirable finite-sample properties when $N=100$. Under the other time periods, it exhibits a severe overrejection problem as well.

\section{Empirical Applications}
\label{sec:empirical_examples}
\subsection{Other price shocks: evidence from existing reduced form applications}

Our model implies that when multiple sectoral prices move together, local exposure to a ``focal'' price is typically correlated with exposure to other prices. In that case, the coefficient on a single price-exposure term conflates several channels unless other price exposures are included explicitly. A few prominent applications already add such controls, but mostly present them as robustness checks. This section documents what those specifications do in \citet{dube2013commodity} and \citet{sviatschi2022making}, and how the magnitudes of their key coefficients change once other price shocks are included. Table~\ref{tab:other-price-shocks} summarizes the main numbers.

\paragraph{Dube and Vargas (2013).}

\citet{dube2013commodity} study how international coffee and oil prices affect civil conflict in Colombian municipalities. Their baseline specification (Table~2) interacts log coffee and log oil prices with pre-determined measures of coffee intensity and oil production, and estimates effects on four measures of violence: guerrilla attacks, paramilitary attacks, clashes, and casualties. For paramilitary attacks, the coefficient on coffee intensity $\times \log p^{\text{coffee}}_t$ is $-0.160$ and the coefficient on oil production $\times \log p^{\text{oil}}_t$ is $0.726$.\footnote{Table~2, column~(2). The coefficients for guerrilla attacks, clashes, and casualties on coffee intensity $\times \log p^{\text{coffee}}_t$ are $-0.611$, $-0.712$, and $-1.828$, respectively.}

In Appendix Table~A.VII, they extend this regression by adding interactions of log prices for other agricultural commodities (sugar, banana, palm, and tobacco) with corresponding pre-determined intensity measures. The coffee coefficient for paramilitary attacks remains negative and highly significant but becomes smaller in magnitude, at $-0.125$. For guerrilla attacks, clashes, and casualties, the coffee coefficients move from $-0.611$, $-0.712$, and $-1.828$ in the baseline to $-0.463$, $-0.639$, and $-1.192$ once sugar, banana, palm, and tobacco price exposures are included. These are non-trivial adjustments---roughly a $20$--$30\%$ reduction in the size of the coffee effect across outcomes---but they leave the qualitative pattern unchanged: higher coffee prices reduce violence in coffee-intensive areas, even after controlling for other commodity price shocks. In the original paper, the multi-commodity specification is presented as a robustness exercise about ``other crops.'' In our framework, it is exactly the sort of specification that recovers a cleaner causal interpretation of the coffee-price exposure coefficient in an environment with multiple relevant price processes.

\paragraph{Sviatschi (2022): child labor.}

\citet{sviatschi2022making} studies how exposure to coca price shocks during childhood affects later outcomes. Her main child-labor specification (Table~I, column~(1)) regresses labor-force participation on a district-level coca price-exposure index, $PriceShock_{dt}$, interacted with age-group indicators. The coefficient on $PriceShock_{dt} \times \mathbf{1}\{11\text{--}14\}$ is $0.144$, capturing the additional increase in child labor at ages 11--14 relative to older ages. The baseline coefficient on $PriceShock_{dt}$ (the effect for the omitted age group) is $0.244$, so the implied total effect at ages 11--14 is $0.244+0.144=0.388$.

Appendix Table~A.II repeats this regression but adds exposure to price shocks for gold, coffee, cacao, cotton, and sugar, each constructed analogously by interacting international prices with pre-determined suitability or deposit measures. Once these additional commodity exposures are included, the coefficient on $PriceShock_{dt} \times \mathbf{1}\{11\text{--}14\}$ remains $0.144$ (with a slightly smaller standard error), while the baseline $PriceShock_{dt}$ coefficient falls from $0.244$ to $0.201$. The implied total effect at ages 11--14 moves from $0.388$ to $0.345$. In contrast, some of the coefficients on other commodities are sizable and significant for child labor. 

\paragraph{Sviatschi (2022): adult incarceration.}

For adult criminal outcomes, \citet{sviatschi2022making} aggregates exposure to coca prices at ages 11--14 into a single index, $PriceShockAge11\text{--}14_{dc}$, and relates it to adult incarceration. In her main specification (Table~III, column~(1)), the coefficient on $PriceShockAge11\text{--}14_{dc}$ for the all-crimes incarceration rate is $3.607$. Table~V then adds exposure to price shocks for gold, coffee, cacao, cotton, and sugar (again interacted with pre-determined suitability or deposits). The coefficient on $PriceShockAge11\text{--}14_{dc}$ declines slightly to $3.523$, while the other commodity exposures are small and statistically indistinguishable from zero. As in the child-labor regressions, the coca exposure coefficient at ages 11--14 is very stable, but the additional commodity shocks matter enough to be worth including if one seeks a structural interpretation of the coca coefficient.

\medskip

Taken together, these examples deliver two messages that our framework clarifies. First, in both applications the focal price-exposure coefficient remains large and significant once other commodity price exposures are controlled for. This is reassuring about the robustness of their substantive conclusions. Second, the fact that the coefficients do move, sometimes by $10$--$30\%$, highlights that omitting other prices changes the object estimated by the focal price-exposure coefficient. In \citet{dube2013commodity} and \citet{sviatschi2022making}, these multi-commodity specifications are presented as robustness checks. Our analysis suggests that, whenever the outcome is plausibly affected by several correlated sectoral prices, such specifications should instead be treated as the preferred empirical implementation of a price-exposure design.

\begin{table}[H]
 \centering \small
 \caption{Price-Exposure Coefficients With and Without Controls for Other Price Shocks}
 \label{tab:other-price-shocks}
 \small \begin{tabular}{lcc}
  \toprule
  & Baseline specification & + Other price shocks \\
  & (1) & (2) \\
  \midrule
  \multicolumn{3}{l}{\textit{Panel A: \citet{dube2013commodity}, Colombia}} \\[0.25em]
  Guerrilla attacks: coffee int.\ $\times \log p^{\text{coffee}}_t$ 
   & $-0.611^{**}$ (0.249) & $-0.463^{***}$ (0.163) \\
  Paramilitary attacks: coffee int.\ $\times \log p^{\text{coffee}}_t$ 
   & $-0.160^{***}$ (0.061) & $-0.125^{***}$ (0.036) \\
  Clashes: coffee int.\ $\times \log p^{\text{coffee}}_t$ 
   & $-0.712^{***}$ (0.246) & $-0.639^{***}$ (0.195) \\
  Casualties: coffee int.\ $\times \log p^{\text{coffee}}_t$ 
   & $-1.828^{*}$ (0.987) & $-1.192^{*}$ (0.639) \\[0.8em]
  \multicolumn{3}{l}{\textit{Panel B: \citet{sviatschi2022making}, Peru}} \\[0.25em]
  Child labor (age 11--14): coca price shock $\times \mathbf{1}\{11\text{--}14\}$ 
   & $0.144^{***}$ (0.028) & $0.144^{***}$ (0.036) \\
  Adult incarceration: coca price shock at ages 11--14 
   & $3.607^{***}$ (0.917) & $3.523^{***}$ (0.901) \\
  \bottomrule
 \end{tabular}
 
 \begin{minipage}{0.9\textwidth}
  \footnotesize
  \medskip
  \footnotesize \noindent \textit{Note:} Each entry reports the coefficient and standard error on the focal price-exposure term in the cited specification, taken directly from the original tables. ${*}$ $p<0.10$, ${**}$ $p<0.05$, ${***}$ $p<0.01$.
  Panel~A uses the coefficients on coffee intensity $\times \log$ coffee price from \citet{dube2013commodity}. 
  Column~(1) reproduces Table~2, columns~(1)--(4), which include exposure to coffee and oil price shocks. 
  Column~(2) uses Appendix Table~A.VII, which augments the regression with exposure to price shocks for sugar, banana, palm, and tobacco. 
  Panel~B uses \citet{sviatschi2022making}. For child labor, column~(1) reports the coefficient on $PriceShock_{dt} \times \mathbf{1}\{11\text{--}14\}$ from Table~I, column~(1); 
  column~(2) reports the same coefficient from Appendix Table~A.II, which adds exposure to price shocks for gold, coffee, cacao, cotton, and sugar. 
  For adult incarceration, column~(1) reports the coefficient on $PriceShockAge11\text{--}14_{dc}$ from Table~III, column~(1), and column~(2) reports the corresponding coefficient from Table~V, column~(1), which adds the same set of other commodity price exposures.
 \end{minipage}
\end{table}

\subsection{Inference in 2SLS: an application to gold mining and violence}
\label{sec:amazon_application}

In this section, we present a stylized empirical application using data on gold mining and homicides in the Brazilian Amazon to compare conventional cluster-robust standard errors with the price–exposure standard errors developed in Section \ref{sec:estimation_inference}. This application is inspired by the setting in \citet{pereira2024tale}, who study the impact of gold market deregulation on violence. Their paper does not use a price–exposure design; instead, they implement a DiD design exploiting a 2013 deregulation law that weakened monitoring of gold transactions. They show that deregulation increased illegal gold mining in protected areas and that municipalities more exposed to illegal mining experienced large increases in homicides after the deregulation.

Because their analysis clearly documents a mechanism linking gold-related rents, illegal mining, and local violence, this setting is a natural context in which to illustrate the implications of using a price–exposure design. In what follows, we implement a deliberately naive price–exposure specification that interacts the number of gold deposits in a municipality with the international gold price. Our goal is purely illustrative: unlike \citet{pereira2024tale}, we abstract from the institutional details of the 2013 reform, omit additional controls, and do not interpret the resulting coefficients as new causal estimates.

We built a panel for 591 municipalities, observed between 2000 and 2022. For each municipality \(i\) and year \(t\), we observe the homicide rate per 100,000 inhabitants, \(Y_{it}\), the area of gold mining within the municipality in square kilometers, \(X_{it}\), and the number of identified gold deposits, \(A_i\).\footnote{For more details about the data and summary statistics, see Appendix \ref{appendix_d}.} We match these data to the international gold price \(p_t\) and construct the standard price–exposure instrument
\[
Z_{it} \equiv (A_i)\log p_t.
\]
 We estimate the canonical price–exposure specification with municipality and year fixed effects. Specifically, we begin with a first-stage regression of mining area on the price–exposure instrument,
\begin{equation*}
 X_{it} = \alpha_i^{X} + \lambda_t^{X} + \pi Z_{it} + v_{it},
 \label{eq:amazon_fs}
\end{equation*}
and a reduced-form regression of homicide rates on the same instrument,
\begin{equation*}
 Y_{it} = \alpha_i^{Y} + \lambda_t^{Y} + \rho Z_{it} + \eta_{it}.
 \label{eq:amazon_rf}
\end{equation*}
Second, we estimate an ordinary least squares (OLS) regression of \(Y_{it}\) on \(X_{it}\),
\begin{equation*}
 Y_{it} = \alpha_i + \lambda_t + \beta^{\text{OLS}} X_{it} + u_{it},
 \label{eq:amazon_ols}
\end{equation*}
and a 2SLS specification that instruments \(X_{it}\) with \(Z_{it}\) in the same two-way fixed-effects framework. With fitted values
\[
 \widehat{X}_{it} = \widehat{\alpha}_i^{X} + \widehat{\lambda}_t^{X} + \widehat{\pi} Z_{it},
\]
the second stage takes the form
\begin{equation*}
 Y_{it} = \alpha_i + \lambda_t + \beta^{\text{2SLS}} \widehat{X}_{it} + \varepsilon_{it}.
 \label{eq:amazon_2sls}
\end{equation*}

All regressions include municipality and year fixed effects and cluster standard errors by municipality. The results are summarized in Table~\ref{tab:amazon_main}. Higher gold prices are strongly associated with increases in mining activity, and the reduced-form and 2SLS estimates suggest that expansions in mining are accompanied by higher homicide rates, consistent with \citet{pereira2024tale}.

\begin{table}[H] \centering \small
 \caption{Gold Prices, Mining, and Homicides in the Brazilian Amazon} 
 \label{tab:amazon_main} 
\begin{tabular}{@{\extracolsep{5pt}}lcccc} 
\hline \\[-1.8ex] 
\\[-1.8ex] & Mining area & \multicolumn{3}{c}{Homicide Rate} \\ 
 & FS: $X$ on $Z$ & RF: $Y$ on $Z$ & OLS: $Y$ on $X$ & 2SLS: $Y$ on $\hat{X}$ \\ 
\\[-1.8ex] & (1) & (2) & (3) & (4)\\ 
\hline \\[-1.8ex] 
 Price exposure $Z_{it}$ & 0.161$^{***}$ & 0.086$^{***}$ & & \\ 
 & (0.023) & (0.026) & & \\ 
 Mining area & & & 0.320$^{***}$ & 0.535$^{**}$ \\ 
 & & & (0.114) & (0.220) \\ 
 \hline \\[-1.8ex] 
Municipality FE & Yes & Yes & Yes & Yes \\ 
Year FE & Yes & Yes & Yes & Yes \\ 
Clusters & Municipality & Municipality & Municipality & Municipality \\ 
Observations & 13,593 & 13,593 & 13,593 & 13,593 \\ 
R$^{2}$ & 0.807 & 0.374 & 0.374 & 0.373 \\ 
\hline 
\hline \\[-1.8ex] 
\end{tabular} \\
\begin{minipage}{0.9\textwidth}
 \medskip
 \justifying % Justify the text
\footnotesize \textit{Note: } This table reports first-stage, reduced-form, OLS, and 2SLS estimates relating gold prices, mining area, and homicide rates in Brazilian Amazon municipalities. All specifications include municipality and year fixed effects. Standard errors, clustered by municipality, are reported in parentheses. See Appendix \ref{appendix_d} for details about the variables. * $p<0.1$; ** $p<0.05$; *** $p<0.01$.
\end{minipage}
\end{table}

We then focus on the 2SLS specification and compare standard errors. Using the same underlying regressions, we compute both the usual cluster-robust variance estimator and the price–exposure variance derived in Section~\ref{sec:estimation_inference}. Table~\ref{tab:amazon_secompare} reports the 2SLS coefficient on mining area together with (i) the municipality-clustered standard error and (ii) the price–exposure standard error. Numerically, the point estimate of \(\beta\) is around \(0.54\): an additional square kilometer of mining area is associated with roughly 0.5 additional homicides per 100,000 inhabitants. The cluster-robust standard error is about \(0.22\), implying statistical significance at conventional levels, whereas the price–exposure standard error is roughly twice as large (about \(0.48\)) and renders the estimate statistically indistinguishable from zero at the 5\% level.

\begin{table}[H] \centering \small
 \caption{2SLS Estimate with Clustered and Price--Exposure Standard Errors} 
 \label{tab:amazon_secompare} 
\begin{tabular}{@{\extracolsep{5pt}}lcc} 
\hline \\[-1.8ex] 
 & \multicolumn{2}{c}{Homicide rate} \\ 
 & Clustered SE & Price--Exposure SE \\ 
\\[-1.8ex] & (1) & (2)\\ 
\hline \\[-1.8ex] 
 Mining area & 0.535$^{**}$ & 0.535 \\ 
 & (0.220) & (0.476) \\ 
 \hline \\[-1.8ex] 
Observations & 13,593 & 13,593 \\ 
R$^{2}$ & 0.373 & 0.373 \\ 
\hline 
\hline \\[-1.8ex] 
\end{tabular} \\
\begin{minipage}{0.9\textwidth}
 \medskip
 \justifying % Justify the text
\footnotesize \textit{Note: } This table reports 2SLS estimates of the effect of mining area on homicide rates in Brazilian Amazon municipalities. Column (1) reports the conventional standard error clustered by municipality; column (2) reports the corresponding price--exposure standard error. See Appendix \ref{appendix_d} for details about the variables. $^{*}$p$<$0.1; $^{**}$p$<$0.05; $^{***}$p$<$0.01.
\end{minipage}
\end{table}

This simple exercise illustrates our main message. A standard empirical implementation of a price–exposure design, using off-the-shelf 2SLS and clustered standard errors, would lead to the conclusion that mining has a precisely estimated effect on violence in this setting. Once we account for the sampling structure implied by the price–exposure design and use the corresponding price–exposure variance, the point estimate is unchanged but the uncertainty around it increases substantially, and the result is no longer statistically significant. Interpreting these estimates substantively would require exactly the kind of careful work done by \citet{pereira2024tale}: tracing the institutional details of deregulation, documenting the mechanisms linking rents to illegal mining and violence, and specifying a research design that aligns with those mechanisms.

\section{Conclusion}
\label{sec:conclusion}
It is a common empirical strategy in applied microeconomics to estimate treatment effects using price-exposure designs. 

In this paper, we develop a multi-sector labor model that microfounds the approach and clarifies what 2SLS and continuous–treatment TWFE estimates recover. Within a finite–population potential–outcomes framework, we show that both estimands can be written as weighted averages of region– and sector–specific causal effects of the focal price, plus contamination terms that depend on causal effects of other sectoral outputs whenever prices co–move or general–equilibrium channels are active. We characterize transparent conditions under which these estimands admit a causal interpretation.

We then study estimation and inference in these designs and quantify the consequences of ignoring the shift–share structure of the residual. We derive a randomization–style asymptotic variance for the 2SLS estimator, show in simulations that conventional Eicker–Huber–White and cluster–robust standard errors can substantially overreject in empirically relevant settings, and propose a feasible price–exposure variance estimator that achieves better coverage. Re–examining prominent applications through this lens, we conclude that price–exposure designs remain powerful tools for applied work, but that credible causal interpretation requires explicit modeling of mechanisms, careful discussion of price co–movement and general–equilibrium forces, and design–consistent inference.

Our results have several implications for applied work. In practice, we recommend that researchers (i) make the theory of change and the associated potential–outcomes model explicit, clarifying which sectoral outputs are allowed to enter the outcome and how the price shock is assumed to operate; (ii) discuss co-movement and general equilibrium channels, and control for other relevant price shocks or sectoral outputs whenever possible, especially in settings where multiple commodities or sectors plausibly matter; (iii) report first stages whenever a measure of sectoral activity is available, and discuss the plausibility of monotonicity and the sources of heterogeneity in the price-output relationship; (iv) be explicit about the validity of exogeneity of the focal price process, which is the central identifying assumption in most settings, rather than parallel trends in outcomes or the exogeneity of exposure shares; and (v) treat short panels with particular caution when conducting inference, given that standard errors that ignore the shift–share structure of residuals can exhibit substantial overrejection in designs with limited time-series variation.

This is still a work in progress. For the next steps, we intend to derive explicit equivalencies to results from previous work, improve the empirical applications, study inference for TWFE and in settings with serial correlation or co-movement in prices, examine implications of adding covariates to the estimation, and expand the labor model to allow for migration and time-varying changes in capital or other structural parameters.

\bibliography{biblio}

% \appendix
% \renewcommand{\thesection}{\Alph{section}} % Sections: A, B, C, ...
% \numberwithin{equation}{section} 

\begin{appendices}

\section{Labor Model Derivation}
\label{appendix_a}
\noindent

In this appendix we verify the expressions used in Section~\ref{sec:economic_model} with an expanded microfoundation and derive the first–stage equation linking tradable–sector output to the exogenous price in Proposition \ref{prop:heterogenous_first_stage}.

\subsection{Microfoundation of sectoral labor demand}

To microfound \eqref{eq:Ld-AKM}, we enrich the environment in Section~\ref{sec:economic_model} with standard assumptions on within‐sector differentiation and price setting, while keeping the regional structure and production technology in \eqref{eq:prod-CD} unchanged.\footnote{This step is identical in spirit to Appendix~C.1 of AKM. We follow similar algebra but adapt all notation to our setting and allow for the additional multiplicative shock $E_{ist}$. Our equation \eqref{eq:Ld-AKM} is analogous to their equation C.4. } Start from a firm producing a differentiated variety
$\omega\ \in \Omega$ in sector $s\in\mathcal S_T$ and region $i$ at time $t$ with Cobb–Douglas technology
\begin{equation}
x_{ist}(\omega)
\;=\;
A_{is}E_{ist}\,k_{ist}(\omega)^{\,1-\theta_s}\,l_{ist}(\omega)^{\,\theta_s},
\qquad 0<\theta_s<1,
\label{eq:variety-prod}
\end{equation}
where $k_{ist}(\omega)$ and $l_{ist}(\omega)$ are capital and labor for variety $\omega$. The sector–region
capital stock is
$K_{is}\equiv\int_{\Omega_s}k_{ist}(\omega)\,d\omega$, which is fixed in the short run.

\paragraph{Unit cost.}
Given wages $w_{it}$ and rental rate $r_{ist}$, a firm that wants to produce one unit of
output solves
\[
\min_{k,l}\; w_{it}l + r_{ist}k
\quad\text{s.t.}\quad
A_{is}E_{ist}k^{1-\theta_s}l^{\theta_s}\ge1.
\]
The Lagrangian
\[
\mathcal L = w_{it}l + r_{ist}k + \lambda\Bigl[1 - A_{is}E_{ist}k^{1-\theta_s}l^{\theta_s}\Bigr]
\]
yields the first–order conditions
\begin{align*}
\partial_l\mathcal L = 0
&\;\Rightarrow\;
w_{it} = \lambda A_{is}E_{ist}\theta_s k^{1-\theta_s}l^{\theta_s-1},\\
\partial_k\mathcal L = 0
&\;\Rightarrow\;
r_{ist} = \lambda A_{is}E_{ist}(1-\theta_s)k^{-\theta_s}l^{\theta_s}.
\end{align*}

Taking the ratio,
\[
\frac{w_{it}}{r_{ist}}
=
\frac{\theta_s}{1-\theta_s}\frac{k}{l}
\quad\Rightarrow\quad
\frac{k}{l}
=
\frac{1-\theta_s}{\theta_s}\frac{w_{it}}{r_{ist}}.
\]

At the sector level (under symmetry of varieties), 
\begin{equation} 
\frac{K_{is}}{L_{ist}} = \iota_s\,\frac{w_{it}}{r_{ist}}, \qquad \iota_s\equiv\frac{1-\theta_s}{\theta_s}>0. 
\label{eq:KL-ratio} \end{equation}

Substituting the optimal factor demands into the cost function
$c_{ist}(w_{it},r_{ist},A_{is}E_{ist})$ gives the dual of
\eqref{eq:variety-prod}:
\begin{equation}
c_{ist}
=
\chi_s\,w_{it}^{\theta_s}r_{ist}^{\,1-\theta_s}(A_{is}E_{ist})^{-1},
\label{eq:unit-cost-der}
\end{equation}
where $\chi_s>0$ depends only on $\theta_s$.

\paragraph{Demand and revenue.}
Within sector $s$, a standard CES aggregator with elasticity of substitution $\sigma_s>1$ delivers isoelastic demand for each differentiated variety. Let $P_{st}$ denote the sectoral CES price index and $C_{st}$ total \emph{nominal} expenditure on sector-$s$ varieties in the world economy. For a variety $\omega$ produced in region $i$,
\[
x_{ist}(\omega)
=
\Bigl(\tfrac{p_{ist}(\omega)}{P_{st}}\Bigr)^{-\sigma_s}C_{st}.
\]
Revenue of variety $\omega$ is
\begin{align}
R_{ist}(\omega)
&\equiv p_{ist}(\omega)\,x_{ist}(\omega)\nonumber\\
&=
\mu_s^{\,1-\sigma_s}\,C_{st}\,P_{st}^{\,\sigma_s-1}\,c_{ist}^{\,1-\sigma_s}\nonumber\\
&=
\tilde\psi_{is}\,
P_{st}^{\,\sigma_s-1}\,
(A_{is}E_{ist})^{\sigma_s-1}\,
w_{it}^{\,\theta_s(1-\sigma_s)}\,
r_{ist}^{\,(1-\theta_s)(1-\sigma_s)},
\label{eq:revenue-der}
\end{align}
where
$
\tilde\psi_{is}
\equiv
\mu_s^{\,1-\sigma_s}C_{st}\chi_s^{\,1-\sigma_s}>0
$
absorbs sectoral constants.
We interpret the economy as a collection of small regions embedded in a large world market.
Sectoral prices $\{P_{st}\}_s$ and nominal expenditures $\{C_{st}\}_s$ are determined in the world economy and are taken as exogenous by each region, so firms treat $(P_{st},C_{st})$ as given when choosing inputs.
Under symmetry of varieties, sectoral revenue is
\[
R_{ist}\;\equiv\;\int_{\Omega_s}R_{ist}(\omega)\,d\omega,
\]
which inherits the same dependence on $(A_{is}E_{ist},w_{it},r_{ist},P_{st})$ up to a constant that we absorb into $\tilde\psi_{is}$.

\paragraph{Labor demand at the variety level.}
With Cobb--Douglas technology, labor's share in variable cost is $\theta_s$.
Total variable cost per variety equals revenue divided by the markup, so
\[
w_{it}l_{ist}(\omega)
=
\theta_s\frac{R_{ist}(\omega)}{\mu_s}
\quad\Rightarrow\quad
l_{ist}(\omega)
=
\underbrace{\frac{\theta_s}{\mu_s}}_{\equiv\upsilon_s}\,
w_{it}^{-1}R_{ist}(\omega).
\]
Under symmetry, sectoral labor and revenue satisfy $L_{ist}\propto l_{ist}(\omega)$ and
$R_{ist}\propto R_{ist}(\omega)$, so we can write
\begin{equation}
L_{ist}
=
\bar\upsilon_s\,w_{it}^{-1}R_{ist},
\label{eq:L-vs-R}
\end{equation}
for some constant $\bar\upsilon_s>0$ that absorbs $\upsilon_s$ and the mass of varieties.

\paragraph{Eliminating capital prices.}
Use \eqref{eq:KL-ratio} to express $r_{ist}$ in terms of $w_{it}$, $L_{ist}$, and $K_{is}$:
\[
\frac{K_{is}}{L_{ist}}
=
\iota_s\,\frac{w_{it}}{r_{ist}}
\quad\Rightarrow\quad
r_{ist}
=
\iota_s\,w_{it}\,\frac{L_{ist}}{K_{is}}.
\]
Substituting into \eqref{eq:revenue-der},
\begin{align}
R_{ist}
&=
\tilde\psi_{is}\,
P_{st}^{\,\sigma_s-1}\,
(A_{is}E_{ist})^{\sigma_s-1}\,
w_{it}^{\,\theta_s(1-\sigma_s)}\,
\biggl[\iota_s\,w_{it}\,\frac{L_{ist}}{K_{is}}\biggr]^{(1-\theta_s)(1-\sigma_s)}
\nonumber\\
&=
\tilde\psi_{is}\,\iota_s^{(1-\theta_s)(1-\sigma_s)}
P_{st}^{\,\sigma_s-1}
(A_{is}E_{ist})^{\sigma_s-1}
w_{it}^{\,\theta_s(1-\sigma_s)+(1-\theta_s)(1-\sigma_s)}
\nonumber\\[-0.25em]
&\hspace{6em}\times
L_{ist}^{\,(1-\theta_s)(1-\sigma_s)}
K_{is}^{\,-(1-\theta_s)(1-\sigma_s)}\nonumber\\
&=
\bar\psi_{is}\,
P_{st}^{\,\sigma_s-1}
(A_{is}E_{ist})^{\sigma_s-1}
w_{it}^{\,1-\sigma_s}
L_{ist}^{\,(1-\theta_s)(1-\sigma_s)}
K_{is}^{\,-(1-\theta_s)(1-\sigma_s)},
\label{eq:R-with-L}
\end{align}
where
$
\bar\psi_{is}
\equiv
\tilde\psi_{is}\,\iota_s^{(1-\theta_s)(1-\sigma_s)}>0
$.

\paragraph{Solving for sectoral labor demand.}
Plugging \eqref{eq:R-with-L} into \eqref{eq:L-vs-R} gives
\begin{align}
L_{ist}
&=
\bar\upsilon_s\,w_{it}^{-1}R_{ist}\nonumber\\
&=
\bar\upsilon_s\bar\psi_{is}
P_{st}^{\,\sigma_s-1}
(A_{is}E_{ist})^{\sigma_s-1}
w_{it}^{-1}
w_{it}^{\,1-\sigma_s}
L_{ist}^{\,(1-\theta_s)(1-\sigma_s)}
K_{is}^{\,-(1-\theta_s)(1-\sigma_s)}\nonumber\\
&=
\tilde\zeta_{is}\,
P_{st}^{\,\sigma_s-1}
(A_{is}E_{ist})^{\sigma_s-1}
w_{it}^{-\sigma_s}
L_{ist}^{\,(1-\theta_s)(1-\sigma_s)}
K_{is}^{\,-(1-\theta_s)(1-\sigma_s)},
\label{eq:L-equation}
\end{align}
where
$
\tilde\zeta_{is}\equiv\bar\upsilon_s\bar\psi_{is}>0
$.

Rearranging \eqref{eq:L-equation}, we isolate the power of $L_{ist}$ on the left:
\begin{align}
L_{ist}
&=
\tilde\zeta_{is}\,
P_{st}^{\sigma_s-1}
(A_{is}E_{ist})^{\sigma_s-1}
w_{it}^{-\sigma_s}
L_{ist}^{(1-\theta_s)(1-\sigma_s)}
K_{is}^{-(1-\theta_s)(1-\sigma_s)}\nonumber\\
\Rightarrow\quad
L_{ist}^{\,1-(1-\theta_s)(1-\sigma_s)}
&=
\tilde\zeta_{is}\,
P_{st}^{\sigma_s-1}
(A_{is}E_{ist})^{\sigma_s-1}
w_{it}^{-\sigma_s}
K_{is}^{-(1-\theta_s)(1-\sigma_s)}.
\label{eq:L-power}
\end{align}
Define
\begin{align}
\Delta_s
&\equiv
1-(1-\theta_s)(1-\sigma_s)\nonumber\\
&=
1-\bigl[1-\theta_s-\sigma_s+\theta_s\sigma_s\bigr]\nonumber\\
&=
\theta_s+\sigma_s-\theta_s\sigma_s\nonumber\\
&=
1+(\sigma_s-1)(1-\theta_s).
\label{eq:Delta-def}
\end{align}
Since $\sigma_s>1$ and $0<\theta_s<1$, we have $\Delta_s>1$, and we set
\begin{equation}
\delta_s
\equiv
\frac{1}{\Delta_s}
=
\frac{1}{1+(\sigma_s-1)(1-\theta_s)}\in(0,1).
\label{eq:delta-def}
\end{equation}
Raising both sides of \eqref{eq:L-power} to the power $\delta_s$ then yields
\begin{align}
L_{ist}
&=
\tilde\zeta_{is}^{\,\delta_s}\,
P_{st}^{\,(\sigma_s-1)\delta_s}
(A_{is}E_{ist})^{\,(\sigma_s-1)\delta_s}
w_{it}^{-\sigma_s\delta_s}
K_{is}^{-(1-\theta_s)(1-\sigma_s)\delta_s}\nonumber\\
&=
\tilde\zeta_{is}^{\,\delta_s}\,
\bigl(A_{is}E_{ist}P_{st}\bigr)^{(\sigma_s-1)\delta_s}
w_{it}^{-\sigma_s\delta_s}
K_{is}^{(1-\theta_s)(\sigma_s-1)\delta_s},
\label{eq:L-full}
\end{align}
where we used $1-\sigma_s=-(\sigma_s-1)$ to rewrite the exponent on $K_{is}$.
Defining
\begin{equation}
\psi_{is}
\equiv
\tilde\zeta_{is}^{\,\delta_s}>0,
\end{equation}
we obtain the sectoral labor demand equation
\begin{equation}
L_{ist}
=
\psi_{is}\;
w_{it}^{-\sigma_s\delta_s}\;
\bigl(A_{is}E_{ist}P_{st}\bigr)^{(\sigma_s-1)\delta_s}\;
K_{is}^{(1-\theta_s)(\sigma_s-1)\delta_s},
\label{eq:Ld-AKM-derived}
\end{equation}
which coincides with \eqref{eq:Ld-AKM}.

\subsection{Proof of Proposition \ref{prop:heterogenous_first_stage}}

\paragraph{Sector-$q$ Employment and Output Responses}

Fix a tradable sector $q\in\mathcal S_T$.
Substituting the wage--price elasticity \eqref{eq:wage-price-elast} into \eqref{eq:dlnL_is_general} with $s=q$ gives the exact log response of sector-$q$ employment to its own price:

\begin{align}
\frac{\partial \ln L_{iqt}}{\partial \ln P_{qt}}
&=
(\sigma_q-1)\delta_q
\;-\;
\sigma_q\delta_q \cdot
\frac{1}{\lambda_i}\sum_{s\in\mathcal S_T}\ell_{is,0}^*(\sigma_s-1)\delta_s\,\rho_{qs}
\label{eq:dlnLiq_dlnPq}
\\
&=
(\sigma_q-1)\delta_q
\left[
1 - 
\frac{\sigma_q}{\sigma_q-1}\cdot\frac{1}{\lambda_i}
\sum_{s\in\mathcal S_T}\ell_{is,0}^*(\sigma_s-1)\delta_s\,\rho_{qs}
\right].
\notag
\end{align}

Sector-$q$ output is, from \eqref{eq:prod-CD},
\begin{equation}
\label{eq:Xiqt}
X_{iqt} \;=\; A_{iq}\,E_{iqt}\,K_{iq}^{\,1-\theta_q}\,L_{iqt}^{\,\theta_q}.
\end{equation}
Taking differentials,
\begin{equation}
\label{eq:dlnXiq}
d\ln X_{iqt} \;=\; d\ln A_{iq} \;+\; d\ln E_{iqt} \;+\; (1-\theta_q)\,d\ln K_{iq} \;+\; \theta_q\,d\ln L_{iqt}.
\end{equation}

Evaluating \eqref{eq:dlnXiq} at a baseline where $A_{iq},K_{iq}$ are fixed (short-run) and focusing on the semi-elasticity w.r.t.\ the exogenous price $P_{qt}$,
\begin{equation}
\label{eq:dlnX_dlnP}
\frac{\partial \ln X_{iqt}}{\partial \ln P_{qt}}
\;=\;
\theta_q\cdot\frac{\partial \ln L_{iqt}}{\partial \ln P_{qt}}
\;=\;
(\sigma_q-1)\theta_q\,\delta_q
\left[
1 - \frac{\sigma_q}{\sigma_q-1}\cdot \frac{1}{\lambda_i}
\sum_{s\in\mathcal S_T}\ell_{is,0}^*(\sigma_s-1)\delta_s\,\rho_{qs}
\right].
\end{equation}

\paragraph{Pre-period equilibrium objects.}
We'll use the semi elasticity to derive a first order expansion of the equilibrium output. For that, we treat baseline objects as fixed and observed in the pre-period:
\[
L_{iq,0}\equiv L_{iq}^*,\qquad
X_{iq,0}\equiv X_{iq}^*,
\]
\[
X_{iq,0}
= A_{iq}\,\bar E_{iq}\,K_{iq}^{\,1-\theta_q}\,(L_{iq,0})^{\theta_q}.
\]
and use this notation below to emphasize they are time-invariant conditioning variables.

Thus, the \emph{levels} semi-elasticity (evaluate at baseline $X_{iq}^*$) is

\begin{equation}
\label{eq:kappa_iq_general}
\tilde\kappa_{iq}
\;\equiv\;
\left.\frac{\partial X_{iqt}}{\partial \ln P_{qt}}\right|_{0}
\;=\; 
X_{iq,0}
\cdot
(\sigma_q-1)\,\theta_q\,\delta_q
\left[
1 - \frac{\sigma_q}{\sigma_q-1}\cdot \frac{1}{\lambda_i}
\sum_{s\in\mathcal S_T}\ell_{is,0}^*(\sigma_s-1)\delta_s\,\rho_{qs}
\right].
\end{equation}

For expositional purposes, also define the scaled elasticity:
\begin{equation}
\label{eq:kappa_scaled}
\kappa_{iq}
\;\equiv\;
\frac{\tilde\kappa_{iq}}{A_{iq}},
\end{equation}

We also need to calculate the semi elasticity of the random shock.
Let $e_{iqt}$ be a centered (log) idiosyncratic efficiency shock defined by
\[
e_{iqt}\;\equiv\;\ln E_{iqt}-\ln\bar E_{iq},
\]
so that
\[
E_{iqt}=\bar E_{iq}\,\exp(e_{iqt}),
\qquad
e_{iqt}=0\ \text{at the baseline }E_{iqt}=\bar E_{iq}.
\]
Since $\partial \ln E_{iqt}/\partial e_{iqt}=1$, the semi elasticities below are invariant to this normalization.
From sectoral labor demand in (\ref{eq:dlnL_is_general}) applied to sector $q$:
\[
d\ln L_{iqt}
\;=\;
(\sigma_q-1)\,\delta_q\,d\ln P_{qt}
\;+\;
(\sigma_q-1)\,\delta_q\,d\ln E_{iqt}
\;-\;
\sigma_q\,\delta_q\,d\ln w_{it},
\]
and holding the common price fixed for this derivative ($d\ln P_{qt}=0$) while $d\ln E_{iqt}=de_{iqt}$, we get
\[
\frac{\partial \ln L_{iqt}}{\partial e_{iqt}}
\;=\;
(\sigma_q-1)\,\delta_q
\;-\;
\sigma_q\,\delta_q\,\frac{\partial \ln w_{it}}{\partial e_{iqt}}.
\]

The market clearing condition in (\ref{eq:MC-lin}) implies the wage semi-elasticity with respect to a \emph{local} shock in $(i,q)$:
\[
\frac{\partial \ln w_{it}}{\partial e_{iqt}}
\;=\;
\frac{1}{\lambda_i}\,\ell^{*}_{iq,0}\,(\sigma_q-1)\,\delta_q,
\qquad
\lambda_i\equiv \phi_i+\sum_{s\in \mathcal S}\ell^{*}_{is,0}\,\sigma_s\,\delta_s,
\]
where $\ell^{*}_{is,0}$ are baseline employment shares. Hence
\[
\frac{\partial \ln L_{iqt}}{\partial e_{iqt}}
\;=\;
(\sigma_q-1)\,\delta_q
\Bigg[
1-\frac{\sigma_q\,\delta_q}{\lambda_i}\,\ell^{*}_{iq,0}
\Bigg].
\]
Using production $X_{iqt}=A_{iq}E_{iqt}K_{iq}^{1-\theta_q}L_{iqt}^{\theta_q}$ gives

\[
\frac{\partial \ln X_{iqt}}{\partial e_{iqt}}
\;=\;
\underbrace{\frac{\partial \ln E_{iqt}}{\partial e_{iqt}}}_{=\,1}
\;+\;
\theta_q\,\frac{\partial \ln L_{iqt}}{\partial e_{iqt}}
\;=\;
1
+
(\sigma_q-1)\,\theta_q\,\delta_q
\Bigg[
1-\frac{\sigma_q\,\delta_q}{\lambda_i}\,\ell^{*}_{iq,0}
\Bigg].
\]
Evaluating at the baseline and multiplying by $X_{iq,0}$ yields
\[
\kappa_{iq}^{E}
\;=\;
\left.\frac{\partial X_{iqt}}{\partial e_{iqt}}\right|_{0}
\;=\;
X_{iq,0}\!\left\{
1
+
(\sigma_q-1)\,\theta_q\,\delta_q
\Bigg[
1-\frac{\sigma_q\,\delta_q}{\lambda_i}\,\ell^{*}_{iq,0}
\Bigg]
\right\}.
\]

\paragraph{First-Order Linear Approximation in Levels}

From Taylor's Theorem, a first-order approximation of \eqref{eq:Xiqt} around the baseline yields the empirical first stage

\begin{equation*}
X_{iqt}
\;\approx\;
X_{iq,0}
\;+\;
\left.\frac{\partial X_{iqt}}{\partial p_{qt}}\right|_{0}\,p_{qt}
\;+\;
\left.\frac{\partial X_{iqt}}{\partial e_{iqt}}\right|_{0}\,e_{iqt}
\end{equation*}

Then, using \eqref{eq:kappa_iq_general}:

\begin{equation}
\label{eq:FS_levels}
 X_{iqt}
\;\approx\;
\alpha_{iq}
\;+\;
\tilde\kappa_{iq}\,p_{qt}
\;+\;
\kappa_{iq}^{E}\,e_{iqt}
\;+\;
\varepsilon_{iqt},
\end{equation}

Define $\kappa_{iq}$ as in \eqref{eq:bar-kappa} so that $\tilde\kappa_{iq}=A_{iq}\,\kappa_{iq}$. Substituting into \eqref{eq:FS_levels} and collecting all shocks orthogonal to $p_{qt}$
(including $\kappa_{iq}^{E}e_{iqt}$) into $\varepsilon_{iqt}$ yields

\begin{equation}
\label{eq:prop-simple-reached}
X_{iqt}
\;\approx\;
\alpha_{iq}
\;+\;
\bigl[A_{iq}\,\kappa_{iq}\bigr]\;p_{qt}
\;+\;
\varepsilon_{iqt},
\end{equation}
which is exactly \eqref{eq:prop-simple}, concluding the proof.

\section{Other Proofs}
\label{appendix_b}
\subsection{Proof of Proposition \ref{prop:beta_2sls}}

\begin{align*}
  &\beta^{2SLS}_{q}=\frac{\sum_{i=1}^{N}\sum_{t=1}^{T}\mathbb{E}\left [ Z_{iqt}Y_{it}|\mathcal{F}_{0} \right ]}{\sum_{i=1}^{N}\sum_{t=1}^{T}\mathbb{E}\left [ Z_{iqt}X_{iqt}|\mathcal{F}_{0} \right ]}=\frac{\sum_{i=1}^{N}\sum_{t=1}^{T}\mathbb{E}\left [ \left ( A_{iq}p_{qt} \right )Y_{it}|\mathcal{F}_{0} \right ]}{\sum_{i=1}^{N}\sum_{t=1}^{T}\mathbb{E}\left [\left ( A_{iq}p_{qt} \right )X_{iqt}|\mathcal{F}_{0} \right ]}\\&=\frac{\sum_{i=1}^{N}\sum_{t=1}^{T}\mathbb{E}\left [ \left ( A_{iq}p_{qt} \right )\left ( \sum_{s=1}^{S}\beta_{is}X_{ist}+\eta_{it} \right )|\mathcal{F}_{0} \right ]}{\sum_{i=1}^{N}\sum_{t=1}^{T}\mathbb{E}\left [\left ( A_{iq}p_{qt} \right )\left ( \alpha_{iq}+\kappa_{iq}\left [ A_{iq}p_{qt} \right ]+\varepsilon_{iqt} \right )|\mathcal{F}_{0} \right ]}\\&=\frac{\sum_{i=1}^{N}\sum_{t=1}^{T}\mathbb{E}\left [ \left ( A_{iq}p_{qt} \right )\left ( \sum_{s=1}^{S}\beta_{is}\left\{ \alpha_{is}+\kappa_{is}\left [ A_{is}p_{st} \right ]+\varepsilon_{ist} \right\}+\eta_{it} \right )|\mathcal{F}_{0} \right ]}{\sum_{i=1}^{N}\sum_{t=1}^{T}\mathbb{E}\left [\left ( A_{iq}p_{qt} \right )\left ( \alpha_{iq}+\kappa_{iq}\left [ A_{iq}p_{qt} \right ]+\varepsilon_{iqt} \right )|\mathcal{F}_{0} \right ]}\\&=\frac{\sum_{i=1}^{N}\sum_{t=1}^{T}\mathbb{E}\left [ \kappa_{iq}\left ( A_{iq}p_{qt} \right )^{2}\beta_{iq}|\mathcal{F}_{0} \right ]}{\sum_{i=1}^{N}\sum_{t=1}^{T}\mathbb{E}\left [ \kappa_{iq}\left ( A_{iq}p_{qt} \right )^{2}|\mathcal{F}_{0} \right ]}+\frac{\sum_{i=1}^{N}\sum_{t=1}^{T}\mathbb{E}\left [ \sum_{s\neq q}\kappa_{is}A_{is}A_{iq}p_{st}p_{qt}\beta_{is} |\mathcal{F}_{0} \right ]}{\sum_{i=1}^{N}\sum_{t=1}^{T}\mathbb{E}\left [ \kappa_{iq}\left ( A_{iq}p_{qt} \right )^{2}|\mathcal{F}_{0} \right ]}\\&=\frac{\sum_{i=1}^{N}\sum_{t=1}^{T} \kappa_{iq}\left ( A_{iq} \right )^{2}\mathbb{V}\left ( p_{qt}|\mathcal{F}_{0} \right )\beta_{iq}}{\sum_{i=1}^{N}\sum_{t=1}^{T} \kappa_{iq}\left ( A_{iq} \right )^{2}\mathbb{V}\left ( p_{qt}|\mathcal{F}_{0} \right )}+\frac{\sum_{i=1}^{N}\sum_{t=1}^{T}\sum_{s\neq q}\kappa_{is}A_{is}A_{iq}\mathbb{C}ov\left ( p_{st},p_{qt}|\mathcal{F}_{0} \right )\beta_{is} }{\sum_{i=1}^{N}\sum_{t=1}^{T} \kappa_{iq}\left ( A_{iq} \right )^{2}\mathbb{V}\left ( p_{qt}|\mathcal{F}_{0} \right )}\\&=\frac{\sum_{i=1}^{N}\sum_{t=1}^{T}\pi_{iqt}\beta_{iq}}{\sum_{i=1}^{N}\sum_{t=1}^{T}\pi_{iqt}}+\frac{\sum_{i=1}^{N}\sum_{t=1}^{T}\left\{ \sum_{s\neq q}\pi_{isqt}\beta_{is}\right\}}{\sum_{i=1}^{N}\sum_{t=1}^{T}\pi_{iqt}}
\end{align*}

\subsection{Proof of Proposition \ref{prop:beta_2sls_general_eq}} 

To obtain this result, we write the numerator of the reduced form as 

\begin{align*}
  &\sum_{i}\sum_{t}\mathbb{E}\left [ Z_{iqt}Y_{it}|\mathcal{\tilde{F}}_{0} \right ]=\sum_{i}\sum_{t}\mathbb{E}\left [ (A_{iq}p_{qt})Y_{it}|\mathcal{\tilde{F}}_{0} \right ]=\sum_{i}\sum_{t}\mathbb{E}\left [ (A_{iq}p_{qt})\left ( \eta_{it}+\sum_{s=1}^{S}X_{ist}\beta_{is} \right )|\mathcal{\tilde{F}}_{0} \right ]\\&=\sum_{i}\sum_{t}\mathbb{E}\left [ (A_{iq}p_{qt})\left ( \eta_{it}+\sum_{s=1}^{S}\left\{ \alpha_{is}+\kappa_{is}(A_{is}p_{st})+\varepsilon_{ist} \right\}\beta_{is} \right )|\mathcal{\tilde{F}}_{0} \right ]\\&\sum_{i}\sum_{t}\tilde{\pi}_{iqt}\beta_{iq}+\sum_{i}\sum_{t}\sum_{s\neq q}\tilde{\pi}_{isqt}\beta_{is}+\sum_{i}\sum_{t}\sum_{s\neq q}\sum_{s'\neq s}A_{iq}\gamma_{iss'}\mathbb{C}ov\left ( p_{qt},p_{s't}|\mathcal{\tilde{F}}_{0} \right )\beta_{is}
\end{align*}

where $\tilde{\pi}_{iqt}=\kappa_{iq}A_{iq}^{2}\mathbb{V}\left(p_{qt}|\mathcal{\tilde{F}}_{0}\right)$ and $\tilde{\pi}_{isqt}=\kappa_{is}A_{is}A_{iq}\mathbb{C}ov\left(p_{qt},p_{st}|\tilde{\mathcal{F}}_{0}\right)$.
\subsection{Proof of Proposition \ref{prop:beta_twfe}}

The proof is similar to the one from Proposition \ref{prop:beta_2sls}, the main difference comes from the decomposition of the denominator. Note that

\begin{equation*}
  \sum_{i=1}^{N}\mathbb{E}\left [ \left ( \Delta Z_{iq} \right )^{2}|\mathcal{F}_{0} \right ]=\sum_{i=1}^{N}\mathbb{E}\left [ \left ( A_{iq}\Delta p_{q} \right )^{2}|\mathcal{F}_{0} \right ]=\sum_{i=1}^{N}A_{iq}^{2}\mathbb{V}\left ( \Delta p_{q}|\mathcal{F}_{0} \right )
\end{equation*}

which concludes the proof.

\section{Asymptotics}
\label{appendix_c}

In this section we provide additional details for the asymptotic results in Section \ref{sec:estimation_inference}.

\subsection{Additional Assumptions}
\begin{assumption}[Regularity Conditions for Consistency]
\label{assumption:regularity_consistency}
\mbox{}\\[-\baselineskip]
\begin{enumerate}
  \item The support of $\beta_{is}$ is bounded.
  \item The second moments of $\eta_{it}$ exist and are uniformly bounded over $i,t$.
  \item The second moments of $\alpha_{is}$ are uniformly bounded over $i,t$.
  \item The second moments of $A_{is}$ are uniformly bounded over $i,t$.
  \item The second moments of $\kappa_{is}$ are uniformly bounded over $i,t$.
  \item $\frac{1}{NT}\sum_{i}\sum_{t}\varepsilon_{ist}\rightarrow 0$, and $\frac{1}{(NT)^{2}}\sum_{i}\sum_{t}\varepsilon_{ist}^{2}\rightarrow 0$.
  \item $\frac{1}{NT}\sum_{i}\sum_{t}\kappa_{is}A_{is}^{2}\mathbb{V}(p_{qs|\mathcal{F}_{0}})$ converges in probability to a strictly positive non-random limit.
  \item For some $\nu>0$, $\mathbb{E}\left [ \left | p_{st}\right |^{2+\nu}|\mathcal{F}_{0} \right ]$ exists and is uniformly bounded.
\end{enumerate}
  
\end{assumption}

\begin{assumption}[Regularity Conditions for Normality]
\label{assumption:regularity_normality}
\mbox{}\\[-\baselineskip]
\begin{enumerate}
  \item The fourth moments of $\eta_{it}$ exist and are uniformly bounded over $i,t$.
  \item The fourth moments of $\alpha_{is}$ are uniformly bounded over $i,t$.
  \item The fourth moments of $A_{is}$ are uniformly bounded over $i,t$.
  \item The fourth moments of $\kappa_{is}$ are uniformly bounded over $i,t$.
  \item For some $\nu>0$, $\mathbb{E}\left [ \left | p_{st}\right |^{4+\nu}|\mathcal{F}_{0} \right ]$ exists and is uniformly bounded.
\end{enumerate}
  
\end{assumption}

\subsection{Proof of Theorem \ref{theorem:asymptotic_variance}}
\subsubsection*{Consistency}
Under Assumptions \ref{assumption:potential_outcomes} and \ref{assumption:regularity_consistency}. Assuming that the price shocks $(p_{q1},...,p_{qT})$ are independent conditional on $\mathcal{F}_{0}$, we can show that 

\begin{equation*}
  \frac{1}{NT}\sum_{i}\sum_{t}Z_{iqt}X_{iqt}=\frac{1}{NT}\sum_{i}\sum_{t}\kappa_{iq}A_{iq}\sigma_{qt}^{2}+o_{p}(1)
\end{equation*}

Begin by writing

\begin{align*}
  &\frac{1}{NT}\sum_{i}\sum_{t}Z_{iqt}X_{iqt}=\frac{1}{NT}\sum_{i}\sum_{t}\left ( A_{iq}p_{qt} \right )\left ( \alpha_{iq}+\kappa(A_{iq}p_{qt})+\varepsilon_{iqt} \right )\\&=\frac{1}{NT}\sum_{i}\sum_{t}\left ( A_{iq}p_{qt} \right )\alpha_{iq}+\frac{1}{NT}\sum_{i}\sum_{t}\kappa_{iq}A_{iq}^{2}(p_{qt}^{2}-\sigma_{qt}^{2})+\frac{1}{NT}\sum_{i}\sum_{t}A_{iq}p_{qt}\varepsilon_{iqt} \\&+\frac{1}{NT}\sum_{i}\sum_{t}\kappa_{iq}A_{iq}\sigma_{qt}^{2}+o_{p}(1)
\end{align*}

The proof follows by showing that the first three terms of the second equality are $o_{p}(1)$.

The first term has mean zero conditional on $\mathcal{F}_{0}$, and variance conditional on $\mathcal{A}_{q}$ equal to

\begin{align*}
  &\mathbb{V}_{\mathcal{A}_{q}}\left ( \frac{1}{NT}\sum_{i}\sum_{t}\left ( A_{iq}p_{qt} \right )\alpha_{iq} \right )=\mathbb{V}_{\mathcal{A}_{q}}\left ( \frac{1}{NT}\sum_{i}A_{iq}\alpha_{iq}\sum_{t} p_{qt} \right )\\&=\frac{1}{(NT)^{2}}\mathbb{E}\left [ \sum_{t}\sigma^{2}_{qt}\sum_{i}(A_{iq}\alpha_{iq})^{2} \right ]\preceq \frac{1}{(NT)^{2}}\mathbb{E}\left [ \sum_{i}(A_{iq}\alpha_{iq})^{2} \right ]\rightarrow 0
\end{align*}

Convergence to zero follows from Assumption \ref{assumption:regularity_consistency}.1 and \ref{assumption:regularity_consistency}.2.

For the second term, note that

\begin{align*}
  &\mathbb{E}\left [ (NT)^{-1}\left | \sum_{t}(p_{qt}^{2}-\sigma_{qt}^{2})\sum_{i}\kappa_{iq}A_{iq}^{2}\right |^{1+\nu/2} \vert \mathcal{F}_{0}\right ]\leq\frac{2}{(NT)^{1+\nu/2}}\sum_{i}(\kappa_{iq}A_{iq}^{2})^{1+\nu/2}\mathbb{E}\left [ \sum_{t}\left | p_{qt}^{2}-\sigma_{qt}^{2}\right |^{1+\nu/2}|\mathcal{F}_{0} \right ]\\&\preceq \frac{1}{(NT)^{1+\nu/2}}\sum_{i}(\kappa_{iq}A_{iq}^{2})^{1+\nu/2}\rightarrow 0
\end{align*}

Where the first inequality follows from Assumption \ref{assumption:regularity_consistency}.6 and the second follows from Assumption \ref{assumption:regularity_consistency}.2 and \ref{assumption:regularity_consistency}.3. The third term has mean zero conditional on $\mathcal{F}_{0}$ and variance

\begin{equation*}
  \mathbb{V}_{\mathcal{A}_{q}}\left ( \frac{1}{NT}\sum_{i}\sum_{t}A_{iq}p_{qt}\varepsilon_{iqt} \right )=\frac{1}{(NT)^{2}}\mathbb{E}_{\mathcal{A}_{q}}\left [ \sum_{i}\sum_{t}\sigma_{qt}^{2}A_{iq}^{2}\varepsilon_{iqt}^{2} \right ]\preceq \frac{1}{(NT)^{2}}\varepsilon_{iqt}^{2}\rightarrow 0 
\end{equation*}

Similarly, we show that

\begin{equation*}
  \frac{1}{NT}\sum_{i}\sum_{t}Z_{iqt}Y_{it}=\frac{1}{NT}\sum_{i}\sum_{t}\kappa_{iq}A_{iq}^{2}\sigma_{qt}^{2}\beta_{iq}+o_{p}(1)
\end{equation*}

First, note that

\begin{align*}
  &\frac{1}{NT}\sum_{i}\sum_{t}Z_{iqt}Y_{it}=\frac{1}{NT}\sum_{i}\sum_{t}(A_{iq}p_{qt})(\eta_{it}+\sum_{s=1}^{s}X_{ist}\beta_{is})\\&=\frac{1}{NT}\sum_{i}\sum_{t}A_{iq}p_{qt}\eta_{it}+\frac{1}{NT}\sum_{i}\sum_{t}(A_{iq}p_{qt})X_{iqt}\beta_{iq}+\frac{1}{NT}\sum_{i}\sum_{t}\sum_{s\neq q}(A_{iq}p_{qt})X_{ist}\beta_{is}
\end{align*}

The first term has mean zero conditional on $\mathcal{F}_{0}$ and variance equal to

\begin{align*}
  &\mathbb{V}_{\mathcal{A}_{q}}\left ( \frac{1}{NT}\sum_{i}\sum_{t}A_{iq}p_{qt}\eta_{it} \right )=\frac{1}{(NT)^{2}}\mathbb{E}_{\mathcal{A}_{q}}\left [ \sum_{i}\sum_{t}\sigma_{qt}^{2}(A_{iq}\eta_{it})^{2} \right ]\\&\preceq\frac{1}{(NT)^{2}}\mathbb{E}_{\mathcal{A}_{q}}\left [ \sum_{i}\sum_{t}(A_{iq}\eta_{it})^{2} \right ]\rightarrow 0
\end{align*}

\subsubsection*{Asymptotic Distribution}

Write the regression error associated to $\beta_{q}$ as 

\begin{align*}
  & u_{it}=Y_{it}-X_{iqt}\beta_{q}=\eta_{it}+X_{iqt}(\beta_{iq}-\beta_{q})+\sum_{s\neq q}X_{ist}\beta_{is}\\&=\eta_{it}+\left\{ \alpha_{iq}+\kappa_{iq}(A_{i}p_{qt})+\varepsilon_{iqt} \right\}(\beta_{iq}-\beta_{q})+\sum_{s\neq q}X_{ist}\beta_{is}\\&\eta_{it}+(\alpha_{iq}+\varepsilon_{iqt})(\beta_{iq}-\beta_{q})+\kappa_{iq}(A_{iq}p_{qt})(\beta_{iq}-\beta_{q})+\sum_{s\neq q}X_{ist}\beta_{is}
\end{align*}

From the consistency result, it follows that

\begin{align*}
  &\widehat{\beta}_{q}-\beta_{q}=(1+o_{p}(1))\left ( \frac{1}{NT}\sum_{i}\sum_{t}\kappa_{iq}A_{iq}^{2}\sigma_{qt}^{2} \right )^{-1}\times\left ( \frac{1}{NT}\sum_{i}\sum_{t}(A_{iq}p_{qt})Y_{it}-\kappa_{iq}A_{iq}^{2}\sigma_{qt}^{2}\beta_{iq} \right )\\&=(1+o_{p}(1))\left ( \frac{1}{NT}\sum_{i}\sum_{t}\kappa_{iq}A_{iq}^{2}\sigma_{qt}^{2} \right )^{-1}\times\left ( \frac{1}{NT}\sum_{i}\sum_{t}(A_{iq}p_{qt})(\eta_{it}+\sum_{s=1}^{S}X_{ist}\beta_{is})-\kappa_{iq}A_{iq}^{2}\sigma_{qt}^{2}\beta_{iq} \right )\\&=(1+o_{p}(1))\left ( \frac{1}{NT}\sum_{i}\sum_{t}\kappa_{iq}A_{iq}^{2}\sigma_{qt}^{2} \right )^{-1}\times\left ( \frac{1}{NT}\sum_{i}\sum_{t}(A_{iq}p_{qt})u_{it}+(A_{iq}p_{qt})X_{iqt}\beta_{q}-\kappa_{iq}A_{iq}^{2}\sigma_{qt}^{2}\beta_{iq} \right )\\&=(1+o_{p}(1))\left ( \frac{1}{NT}\sum_{i}\sum_{t}\kappa_{iq}A_{iq}^{2}\sigma_{qt}^{2} \right )^{-1}\\&\times\left ( \frac{1}{NT}\sum_{i}\sum_{t}(A_{iq}p_{qt})u_{it}+(A_{iq}p_{qt})(\alpha_{iq}+\varepsilon_{iqt} )\beta_{q}+\kappa_{iq}A_{iq}^{2}p_{qt}^{2}\beta_{q}-\kappa_{iq}A_{iq}^{2}\sigma_{qt}^{2}\beta_{iq} \right )\\&=(1+o_{p}(1))\left ( \frac{1}{NT}\sum_{i}\sum_{t}\kappa_{iq}A_{iq}^{2}\sigma_{qt}^{2} \right )^{-1}\\&\times \left ( \frac{1}{NT}\sum_{i}\sum_{t}(A_{iq}p_{qt})u_{it}+(A_{iq}p_{qt})(\alpha_{iq}+\varepsilon_{iqt} )\beta_{q}+\kappa_{iq}A_{iq}^{2}p_{qt}^{2}\beta_{q}-\kappa_{iq}A_{iq}^{2}\sigma_{qt}^{2}(\beta_{iq}-\beta_{q})-\kappa_{iq}A_{iq}^{2}\sigma_{qt}^{2}\beta_{q} \right )
\end{align*}

Note that $\frac{1}{NT}\sum_{i}\sum_{t}\kappa_{iq}A_{iq}^{2}\sigma_{qt}^{2}(\beta_{iq}-\beta_{q})=0$ and hence, it follows that 

\begin{align*}
  &\widehat{\beta}_{q}-\beta_{q}=(1+o_{p}(1))\left ( \frac{1}{NT}\sum_{i}\sum_{t}\kappa_{iq}A_{iq}^{2}\sigma_{qt}^{2} \right )^{-1}\\&\times \left ( \frac{1}{NT}\sum_{i}\sum_{t}(A_{iq}p_{qt})u_{it}+(A_{iq}p_{qt})(\alpha_{iq}+\varepsilon_{iqt} )\beta_{q}+\kappa_{iq}A_{iq}^{2}(p_{qt}^{2}-\sigma_{qt}^{2} )\beta_{q} \right )
\end{align*}

And therefore, 

\begin{align*}
  &\sqrt{NT}(\widehat{\beta}_{q}-\beta_{q})=(1+o_{p}(1))\left ( \frac{1}{NT}\sum_{i}\sum_{t}\kappa_{iq}A_{iq}^{2}\sigma_{qt}^{2} \right )^{-1}\\&\times \left ( \frac{1}{\sqrt{NT}}\sum_{i}\sum_{t}(A_{iq}p_{qt})u_{it}+(A_{iq}p_{qt})(\alpha_{iq}+\varepsilon_{iqt} )\beta_{q}+\kappa_{iq}A_{iq}^{2}(p_{qt}^{2}-\sigma_{qt}^{2} )\beta_{q} \right )\\&=(1+o_{p}(1))\left ( \frac{1}{NT}\sum_{i}\sum_{t}\kappa_{iq}A_{iq}^{2}\sigma_{qt}^{2} \right )^{-1}\times \left ( \frac{1}{\sqrt{NT}}\sum_{i}\sum_{t}(A_{iq}p_{qt})u_{it}+o_{p}(1) \right )
\end{align*}

In order to prove the result, it suffices to show that $\frac{1}{\sqrt{NT}}\sum_{i}\sum_{t}(A_{iq}p_{qt})u_{it}\overset{d}{\rightarrow}N(0,V_{NT})$. We do so by showing that the Lindeberg conditions hold, and thus the convergence in distribution follows from the Lindeberg CLT.

We decompose $\frac{1}{\sqrt{NT}}\sum_{i}\sum_{t}(A_{iq}p_{qt})u_{it}$ as 

\begin{align*}
  &\frac{1}{\sqrt{NT}}\sum_{i}\sum_{t}(A_{iq}p_{qt})u_{it}=\frac{1}{\sqrt{NT}}\sum_{i}\sum_{t}(A_{iq}p_{qt})\eta_{it}+\frac{1}{\sqrt{NT}}\sum_{i}\sum_{t}(A_{iq}p_{qt})(\alpha_{iq}+\varepsilon_{ist})(\beta_{iq}-\beta_{q})\\&+\frac{1}{\sqrt{NT}}\sum_{i}\sum_{t}\kappa_{iq}A_{iq}^{2}p_{qt}^{2}(\beta_{iq}-\beta_{q})+\frac{1}{\sqrt{NT}}\sum_{i}\sum_{t}\sum_{s\neq q}(A_{iq}p_{qt})(\alpha_{is}+\varepsilon_{ist})\beta_{is}\\&+\frac{1}{\sqrt{NT}}\sum_{i}\sum_{t}\sum_{s\neq q}\kappa_{is}A_{is}A_{is}p_{st}p_{qt}\beta_{is}\\&= \frac{1}{\sqrt{NT}}\sum_{i}\sum_{t}(A_{iq}p_{qt})\eta_{it}+\frac{1}{\sqrt{NT}}\sum_{i}\sum_{t}(A_{iq}p_{qt})(\alpha_{iq}+\varepsilon_{ist})(\beta_{iq}-\beta_{q})\\&+\frac{1}{\sqrt{NT}}\sum_{i}\sum_{t}\kappa_{iq}A_{iq}^{2}(p_{qt}^{2}-\sigma_{qt}^{2})(\beta_{iq}-\beta_{q})+\frac{1}{\sqrt{NT}}\sum_{i}\sum_{t}\sum_{s\neq q}(A_{iq}p_{qt})(\alpha_{is}+\varepsilon_{ist})\beta_{is}\\&+\frac{1}{\sqrt{NT}}\sum_{i}\sum_{t}\sum_{s\neq q}\kappa_{is}A_{is}A_{is}p_{st}p_{qt}\beta_{is}=\frac{1}{NT}\sum_{i}\sum_{t}\mathcal{Y}_{it}^{(1)}+\mathcal{Y}_{it}^{(2)}+\mathcal{Y}_{it}^{(3)}+\mathcal{Y}_{it}^{(4)}+\mathcal{Y}_{it}^{(5)}\\&=\frac{1}{NT}\sum_{i}\sum_{t}\mathcal{Y}_{it}
\end{align*}

To show that the Lindeberg condition holds, it suffices to show that for some $\nu>0$, $\frac{1}{(NT)^{1+\nu/4}}\sum_{i}\sum_{t}\mathbb{E}_{\mathcal{A}_{q}}\left [ \mathcal{Y}_{it}^{2+\nu/2} \right ]\rightarrow 0$.

The Lindeberg condition holds if 

\begin{align*}
  &\frac{1}{(NT)^{2}}\sum_{i}\sum_{t}\mathbb{E}_{\mathcal{A}_{q}}\left [ \left ( \mathcal{Y}_{it}^{(1)} \right )^{4} \right ]\rightarrow 0,\ \frac{1}{(NT)^{2}}\sum_{i}\sum_{t}\mathbb{E}_{\mathcal{A}_{q}}\left [ \left ( \mathcal{Y}_{it}^{(2)} \right )^{4} \right ]\rightarrow 0,\\&\frac{1}{(NT)^{1+\nu/4}}\sum_{i}\sum_{t}\mathbb{E}_{\mathcal{A}_{q}}\left [ \left ( \mathcal{Y}_{it}^{(3)} \right )^{2+\nu/2} \right ]\rightarrow 0, \ \frac{1}{(NT)^{2}}\sum_{i}\sum_{t}\mathbb{E}_{\mathcal{A}_{q}}\left [ \left ( \mathcal{Y}_{it}^{(4)} \right )^{4} \right ]\rightarrow 0,\\& \frac{1}{(NT)^{2}}\sum_{i}\sum_{t}\mathbb{E}_{\mathcal{A}_{q}}\left [ \left ( \mathcal{Y}_{it}^{(5)} \right )^{4} \right ]\rightarrow 0
\end{align*}

The first follows from Assumptions \ref{assumption:regularity_normality}.1 and \ref{assumption:regularity_normality}.\ref{assumption:regularity_consistency}. The second follows from Assumptions \ref{assumption:regularity_consistency}.1, \ref{assumption:regularity_consistency}.6, \ref{assumption:regularity_normality}.1 and \ref{assumption:regularity_normality}.\ref{assumption:regularity_consistency}. The third follows from Assumptions \ref{assumption:regularity_normality}.3, \ref{assumption:regularity_normality}.4 and \ref{assumption:regularity_normality}.5. The fourth follows from Assumptions \ref{assumption:regularity_consistency}.1, \ref{assumption:regularity_consistency}.6, \ref{assumption:regularity_normality}.2 and \ref{assumption:regularity_normality}.5. The fifth follows from Assumptions \ref{assumption:regularity_consistency}.1, \ref{assumption:regularity_normality}.2 and \ref{assumption:regularity_normality}.\ref{assumption:regularity_normality}. Therefore, it follows from the Lindeberg CLT that

\begin{equation*}
  \frac{1}{\sqrt{NT}}\sum_{i}\sum_{t}(A_{iq}p_{qt})u_{it}\overset{d}{\rightarrow}N(0,V_{NT})
\end{equation*}

which concludes the proof.

\subsubsection*{Inference under Heterogeneous Effects}

For valid, yet conservative, inference under heterogeneous effects, it must be the case that

\begin{equation*}
  \frac{\sum_{t}p_{qt}^{2}\widehat{R}_{t}^{2}}{NT}\geq\mathcal{V}_{NT}+o_{p}(1)
\end{equation*}

In order to assess the conditions under which the inequality holds, note that

\begin{align*}
  &\mathcal{V}_{NT}=\frac{\mathbb{V}\left( \sum_{t}p_{qt}R_{t}|\mathcal{F}_{0} \right)}{NT}=\frac{\sum_{t}\mathbb{E}\left[ p_{qt}^{2}R_{t}^{2}|\mathcal{F}_{0} \right]}{NT}+\frac{\sum_{t\neq t'}\mathbb{C}ov\left( p_{qt}R_{t},p_{qt'}R_{t'}|\mathcal{F}_{0} \right)}{NT}-\frac{\sum_{t}\mathbb{E}\left[ p_{qt}R_{t}|\mathcal{F}_{0} \right]^{2}}{NT}\\&=D_{1}+D_{2}-D_{3}
\end{align*}

Note that 

\begin{align*}
  &D_{2}=\frac{\sum_{t\neq t'}\sigma_{qt}^{2}\sigma_{qt'}^{2}\sum_{i,j}\kappa_{iq}^{2}A_{iq}^{4}(\beta_{iq}-\beta)\kappa_{jq}^{2}A_{jq}^{4}(\beta_{jq}-\beta_{q})}{NT},\\& D_{3}=\frac{\sum_{t}\left( \sum_{i}\kappa_{iq}A_{iq}^{2}\sigma_{qt}^{2}\left( \beta_{iq}-\beta_{q} \right) \right)^{2}}{NT}
\end{align*}

And thus, in the case of homogeneous effects $D_{2}=D_{3}=0$, the standard error estimator consistently estimates $D_{1}$. In order for inference to be valid, yet conservative, under treatment effect heterogeneity, it must be the case that $D_{2}=o_{p}(1)$.

\section{Empirical application details}
\label{appendix_d}
All data used in the empirical application come from publicly available sources: 
DATASUS\footnote{\url{https://datasus.saude.gov.br/}} (mortality microdata from SIM, considering ICD-10 codes X86-Y09), Serviço Geológico do Brasil\footnote{\url{https://geoportal.cprm.gov.br/geosgb/}} (mineral deposit locations), IBGE\footnote{\url{https://www.ibge.gov.br/estatisticas/economicas/contas-nacionais.html}} (municipal population), World Bank\footnote{\url{https://www.worldbank.org/en/research/commodity-markets}} (international gold prices), and MapBiomas\footnote{\url{https://mapbiomas.org/en/download-dos-atbds?cama_set_language=en}} (machine-learning based satellite mining data). 

\begin{figure}[H]
 \centering
 \caption{Gold Prices and Mining Area in the Brazilian Amazon}
 \label{fig:amazon_price_mining}

 \begin{subfigure}{0.5\textwidth}
  \centering
  \includegraphics[width=\textwidth]{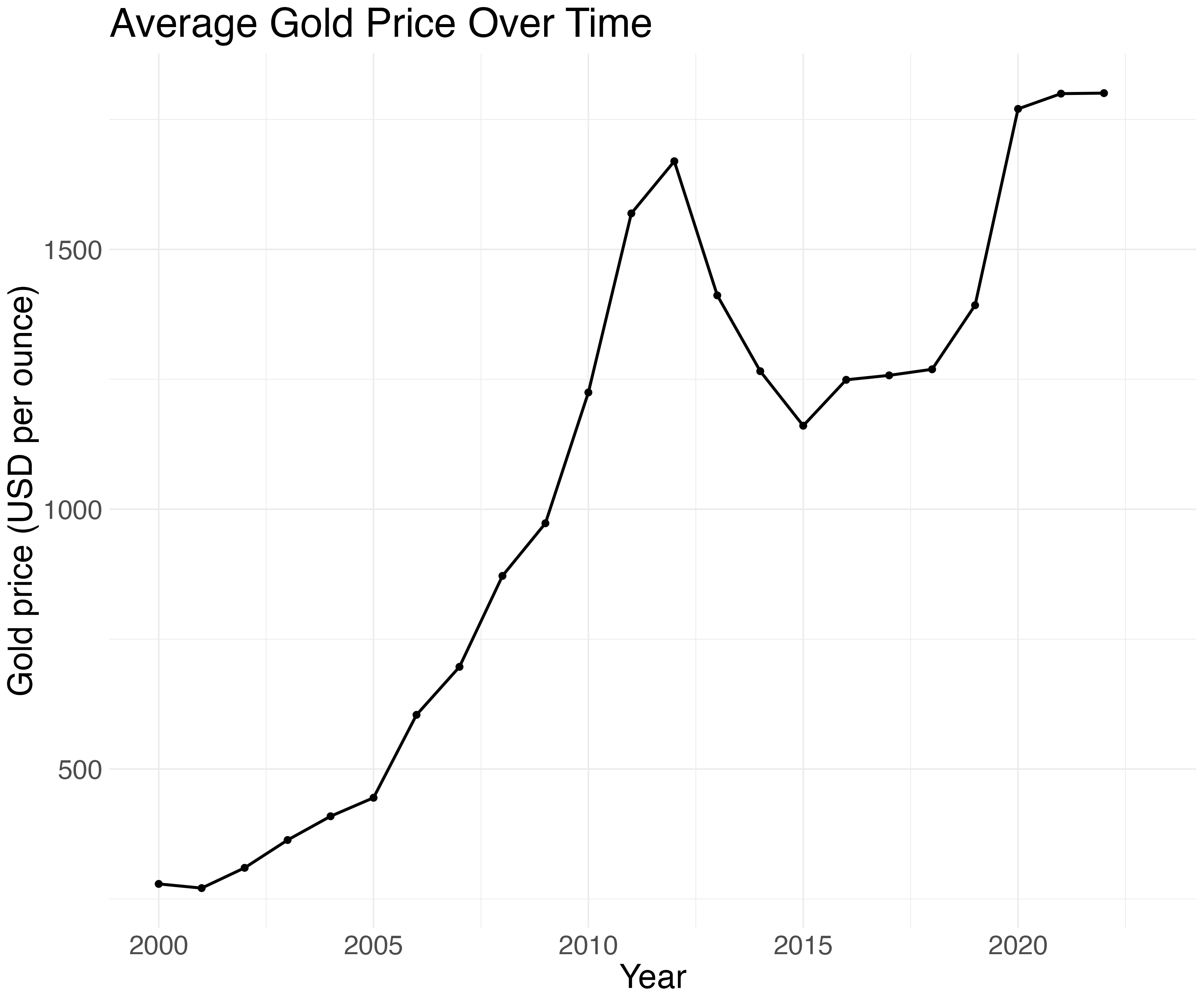}
  \caption{Average gold price over time.}
  \label{fig:amazon_price}
 \end{subfigure}

 \vspace{0.75em}

 \begin{subfigure}{0.5\textwidth}
  \centering
  \includegraphics[width=\textwidth]{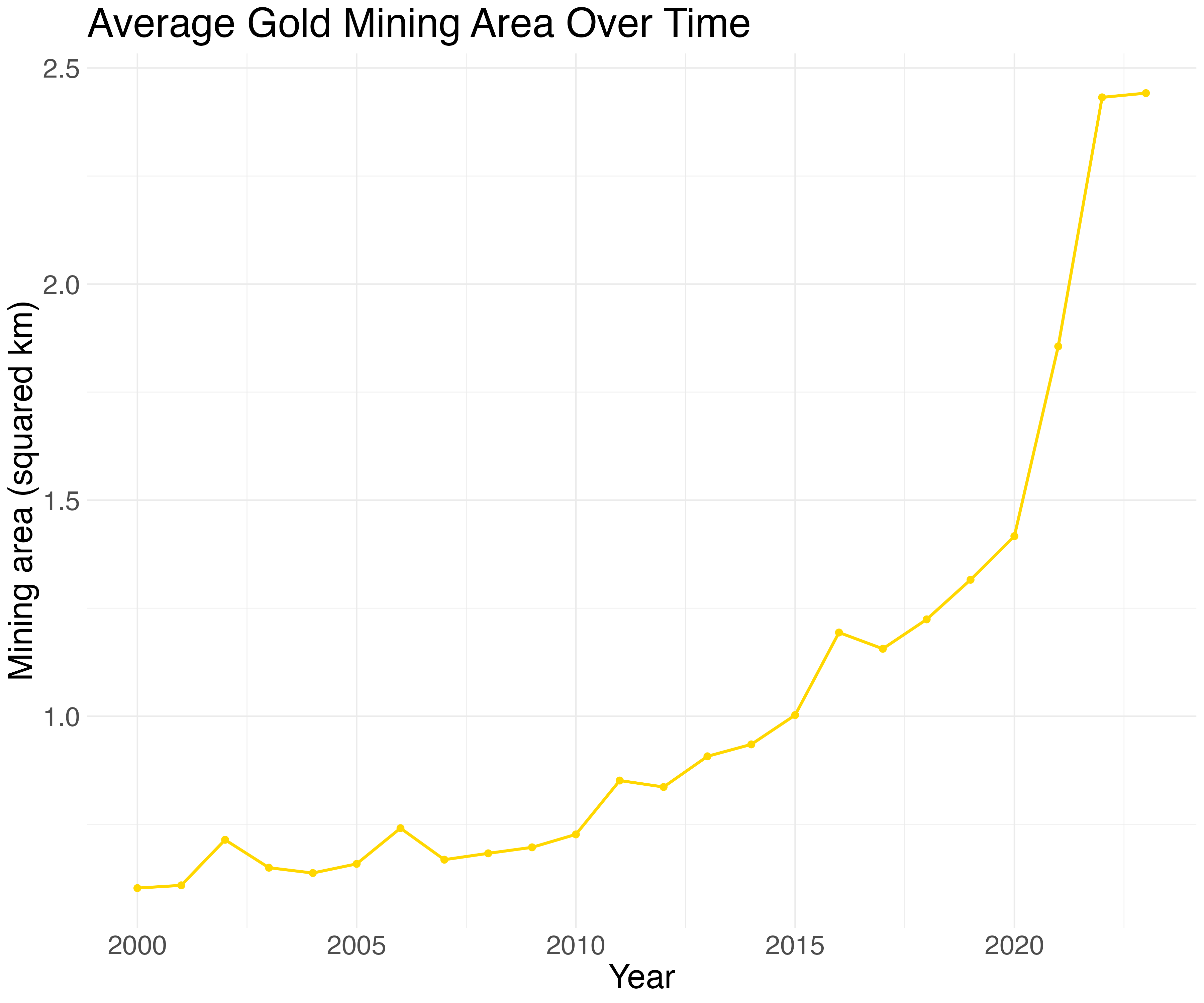}
  \caption{Average gold mining area over time.}
  \label{fig:amazon_mining}
 \end{subfigure}

 \vspace{0.5em}
 \footnotesize \noindent \textit{Note:} Panel (a) plots the average international gold price (USD per ounce) by year. Panel (b) plots the average gold mining area (km$^2$) across municipalities by year. Both series are computed from the same municipality--year panel used in the empirical application.
\end{figure}

\begin{figure}[H]
 \centering
 \caption{Homicides and Gold Deposit Intensity in the Brazilian Amazon}
 \label{fig:amazon_homicides}
 \includegraphics[width=0.7\textwidth]{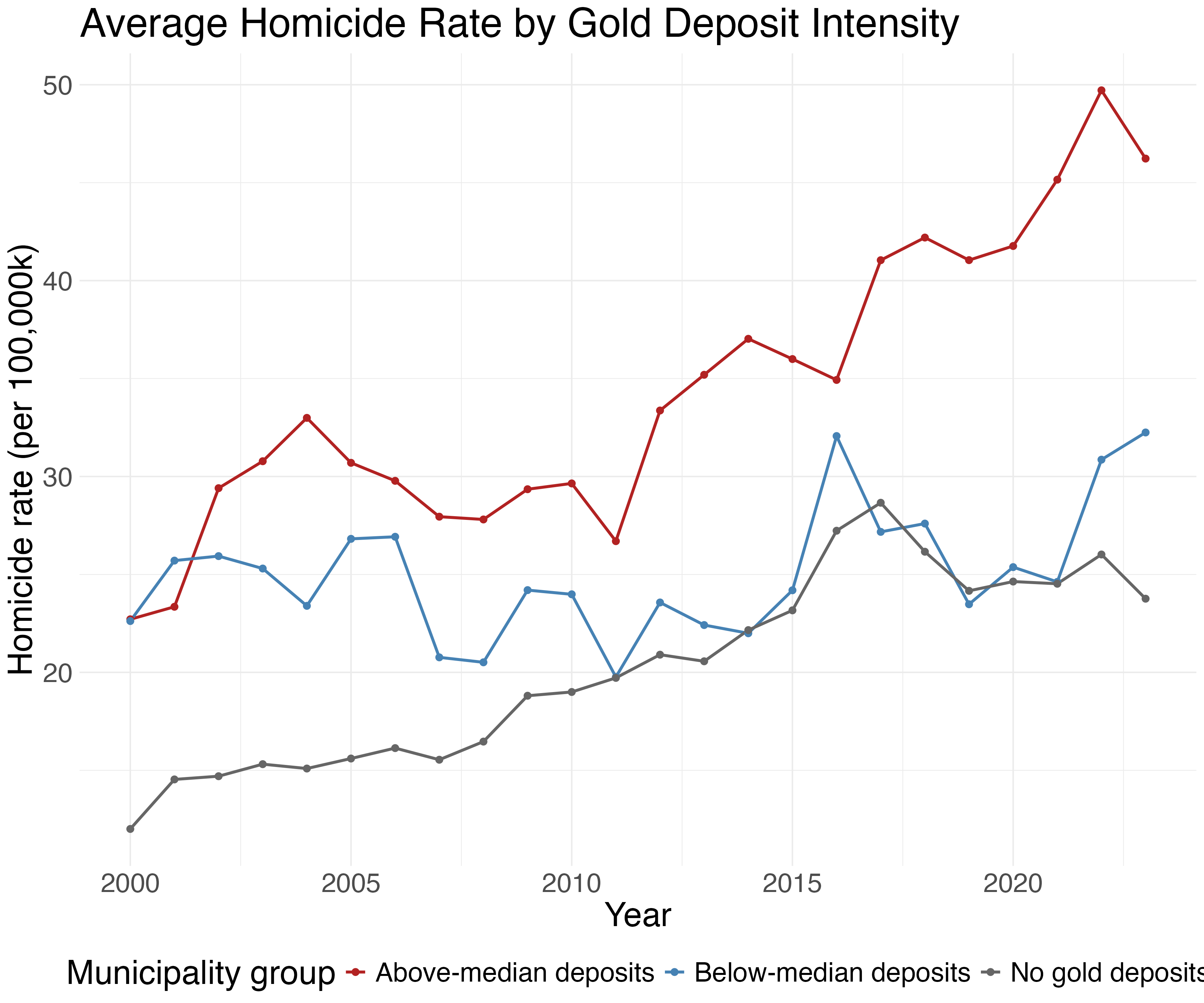}

 \vspace{0.5em}
 \footnotesize \noindent \textit{Note:} The figure plots average homicide rates per 100{,}000 inhabitants by municipality group, where groups are defined by the number of gold deposits (no deposits, below-median deposits, and above-median deposits among municipalities with positive deposits). Series are computed from the municipality--year panel used in the empirical application.
\end{figure}

There are 591 municipalities in the panel, observed between 2000 and 2022. Table \ref{tab:amazon_sumstats} reports summary statistics for the variables used. The average homicide rate is about 22 per 100,000 inhabitants, with substantial cross-sectional variation and a long right tail. Mining activity is highly concentrated: mean mining area is roughly \(1\)~km\(^2\), with most municipality–years at zero and a small number of locations with very large mines. The count of deposits and the price-exposure variable \(Z_{it}\) display similarly skewed distributions. Figures \ref{fig:amazon_price_mining} and \ref{fig:amazon_homicides} show the evolution of gold prices, mining area, and homicide rates over time, illustrating the rise in gold prices was accompanied by an increase in mining activity (which motivates the first stage) and in homicides in municipalities with more gold deposits (which motivates the empirical analysis).

\begin{table}[!htbp] \centering 
 \caption{Summary Statistics} 
 \label{tab:amazon_sumstats} 
\begin{tabular}{@{\extracolsep{5pt}}lccccc} 
\\[-1.8ex]\hline 
\hline \\[-1.8ex] 
Statistic & \multicolumn{1}{c}{N} & \multicolumn{1}{c}{Mean} & \multicolumn{1}{c}{St. Dev.} & \multicolumn{1}{c}{Min} & \multicolumn{1}{c}{Max} \\ 
\hline \\[-1.8ex] 
Homicide rate (per 100k) & 13,593 & 22.38 & 27.47 & 0.00 & 658.32 \\ 
Mining area (km$^2$) & 13,593 & 0.98 & 7.69 & 0.00 & 412.22 \\ 
Gold deposits (count) & 13,593 & 3.92 & 21.29 & 0 & 459 \\ 
Gold price (USD/oz) & 13,593 & 1,046.22 & 515.99 & 270.99 & 1,800.60 \\ 
Price exposure Z & 13,593 & 26.61 & 145.10 & 0.00 & 3,440.61 \\ 
\hline \\[-1.8ex] 
\end{tabular}
\begin{minipage}{0.9\textwidth}
 \medskip
 \justifying % Justify the text
\footnotesize \textit{Note: } This table reports summary statistics for the variables used in Tables \ref{tab:amazon_main} and \ref{tab:amazon_secompare}. Statistics are calculated over the full panel of 591 municipalities and 23 years.
\end{minipage}
\end{table}

\end{appendices}
\end{document}